\documentclass[lettersize,journal]{IEEEtran}
\usepackage{amsmath,amsfonts}
\usepackage{amssymb}
\usepackage{amsthm}
\usepackage{algorithm}
\usepackage{array}
\usepackage[font={footnotesize}]{caption}
\usepackage{subcaption}
\usepackage{textcomp}
\usepackage{stfloats}
\usepackage{url}
\usepackage{verbatim}
\usepackage{balance}
\usepackage{graphicx}
\usepackage{cite}
\usepackage{mathtools}
\usepackage[nolist]{acronym}
\usepackage{bm}
\usepackage[mode=image]{standalone}
\usepackage{tikz}
\usepackage{xcolor}
\usepackage{multirow}
\usepackage{pgfplots}
\usepackage{bigfoot} %This is needed to force one of the footnotes to be on the same page as the text
\usepackage{dsfont} %To do \mathds{1} as in \mathbb{1}
\usepackage{booktabs}
\usetikzlibrary{arrows.meta}
\tikzset{every picture/.style={line width=0.6pt}}
\usetikzlibrary{arrows.meta,
                calc, 
                backgrounds,
                chains,
                fit,
                quotes,
                shapes.geometric,
                positioning,
                intersections,
                spath3}

\usepgfplotslibrary{colormaps} % LATEX and plain TEX
\usetikzlibrary{pgfplots.colormaps} % LATEX and plain TEX

\usepackage{algpseudocode}
\usepackage{microtype}

\newtheorem{proposition}{Proposition}
\newtheorem{lemma}{Lemma}
\newtheorem{remark}{Remark}

\DeclareDocumentCommand{\complexset}{o o}{%
    \mathbb{C}\IfValueT{#1}{\IfValueTF{#2}{^{#1\times#2}}{^{#1}}}
}
\DeclareDocumentCommand{\realset}{o o}{%
    \mathbb{R}\IfValueT{#1}{\IfValueTF{#2}{^{#1\times#2}}{^{#1}}}
}

\DeclareDocumentCommand{\abm}{o}{%
    \bm{\mathrm{a}}\IfValueT{#1}{^{(#1)}}
}

\DeclareDocumentCommand{\cbm}{o}{%
    \bm{\mathrm{c}}\IfValueT{#1}{^{(#1)}}
}

\DeclareDocumentCommand{\fbm}{o o}{%
    \bm{\mathrm{f}}\IfValueT{#1}{\IfValueTF{#2}{^{(#1,#2)}}{^{(#1)}}}
}

\DeclareDocumentCommand{\Hbm}{o o}{%
    \bm{\mathrm{H}}\IfValueT{#1}{\IfValueTF{#2}{^{(#1,#2)}}{^{(#1)}}}
}
\DeclareDocumentCommand{\hbm}{o o}{%
    \bm{\mathrm{h}}\IfValueT{#1}{\IfValueTF{#2}{^{(#1,#2)}}{^{(#1)}}}
}

\DeclareDocumentCommand{\hbmbar}{o o}{%
    \bm{\bar{\mathrm{h}}}\IfValueT{#1}{\IfValueTF{#2}{^{(#1,#2)}}{^{(#1)}}}
}

\DeclareDocumentCommand{\Hbmbar}{o}{%
    \bm{\bar{\mathrm{H}}}\IfValueT{#1}{^{(#1)}}
}

\DeclareDocumentCommand{\Hbmtilde}{o o}{%
    \tilde{\bm{\mathrm{H}}}\IfValueT{#1}{\IfValueTF{#2}{^{(#1,#2)}}{^{(#1)}}}
}

\DeclareDocumentCommand{\Lfrak}{o}{%
    \mathfrak{L}\IfValueT{#1}{_{\mathrm{#1}}}
}

\DeclareDocumentCommand{\pbm}{o o}{%
    \bm{\mathrm{p}_{\theta}}\IfValueT{#1}{\IfValueTF{#2}{^{(#1,#2)}}{^{(#1)}}}
}
\DeclareDocumentCommand{\pbmi}{o o}{%
    \bm{\mathrm{p}_{\theta^{(i)}}}\IfValueT{#1}{\IfValueTF{#2}{^{(#1,#2)}}{^{(#1)}}}
}

\DeclareDocumentCommand{\pbmstar}{o}{%
    \bm{\mathrm{p}_{\theta^{\star}}}\IfValueT{#1}{^{(#1)}}
}

\DeclareDocumentCommand{\vbm}{o o}{%
    \bm{\mathrm{v}}\IfValueT{#1}{\IfValueTF{#2}{^{(#1,#2)}}{^{(#1)}}}
}
\DeclareDocumentCommand{\Vbmbar}{o o}{%
    \bar{\bm{\mathrm{V}}}\IfValueT{#1}{\IfValueTF{#2}{^{(#1,#2)}}{^{(#1)}}}
}
\DeclareDocumentCommand{\vbmtilde}{o o}{%
    \tilde{\bm{\mathrm{v}}}_{\thetabm}\IfValueT{#1}{\IfValueTF{#2}{^{(#1,#2)}}{^{(#1)}}}
}
\DeclareDocumentCommand{\vbmtildei}{o o}{%
    \tilde{\bm{\mathrm{v}}}_{\thetabm^{(i)}}\IfValueT{#1}{\IfValueTF{#2}{^{(#1,#2)}}{^{(#1)}}}
}
\DeclareDocumentCommand{\vbmtildestar}{o}{%
    \tilde{\bm{\mathrm{v}}}_{\thetabm^{\star}}\IfValueT{#1}{^{(#1)}}
}
\DeclareDocumentCommand{\vbmbar}{o o}{%
    \bar{\bm{\mathrm{v}}}\IfValueT{#1}{\IfValueTF{#2}{^{(#1,#2)}}{^{(#1)}}}
}

\DeclareDocumentCommand{\xbm}{o o}{%
    \bm{\mathrm{x}}\IfValueT{#1}{\IfValueTF{#2}{^{(#1,#2)}}{^{(#1)}}}
}

\DeclareDocumentCommand{\xbmunder}{o o}{%
    \bm{\underline{\mathrm{x}}}\IfValueT{#1}{\IfValueTF{#2}{^{(#1,#2)}}{^{(#1)}}}
}

\DeclareDocumentCommand{\xbmhat}{o o}{%
    \hat{\bm{\mathrm{x}}}_{\bm{\theta}}\IfValueT{#1}{\IfValueTF{#2}{^{(#1,#2)}}{^{(#1)}}}
}

\DeclareDocumentCommand{\xbmhati}{o o}{%
    \hat{\bm{\mathrm{x}}}_{\bm{\theta}^{(i)}}\IfValueT{#1}{\IfValueTF{#2}{^{(#1,#2)}}{^{(#1)}}}
}

\DeclareDocumentCommand{\xbmhatstar}{o}{%
    \hat{\bm{\mathrm{x}}}_{\bm{\theta}^{\star}}\IfValueT{#1}{^{(#1)}}
}

\DeclareDocumentCommand{\wbmbar}{o }{%
    \bar{\bm{\mathrm{w}}}\IfValueT{#1}{^{(#1)}}
}

\DeclareDocumentCommand{\ubmbar}{o }{%
    \bar{\bm{\mathrm{u}}}\IfValueT{#1}{^{(#1)}}
}

\DeclareDocumentCommand{\Cset}{o }{%
    \mathcal{C}\IfValueT{#1}{^{(#1)}}
}

\DeclareDocumentCommand{\Lset}{o}{%
    \mathcal{L}\IfValueT{#1}{^{(#1)}}
}

\newcommand{\Aset}{\mathcal{A}}
\newcommand{\Abm}{\bm{\mathrm{A}}}

\newcommand{\Bset}{\mathcal{B}}
\newcommand{\Bbm}{\bm{\mathrm{B}}}
\newcommand{\bbm}{\bm{\mathrm{b}}}

\newcommand{\CT}{\mathrm{CT}}

\newcommand{\Dh}{D_{\mathrm{h}}}
\newcommand{\Dv}{D_{\mathrm{v}}}
\newcommand{\Dx}{D_{\mathrm{x}}}
\newcommand{\dft}[1]{\bm{\mathrm{F}}_{#1}}
\newcommand{\dhat}{\hat{d}}

\newcommand{\gtheta}[1]{g_{\bm{\theta}}(#1)}

\newcommand{\hermit}{\mathsf{H}}

\newcommand{\Jsetn}{\mathcal{J}^{(n)}}
\newcommand{\Jsethatn}{\hat{\mathcal{J}}^{(n)}}

\newcommand{\KS}{\mathrm{KS}}

\newcommand{\lambdaCC}{\lambda_{\mathrm{CC}}}
\newcommand{\lambdadt}{\lambda_{\mathrm{DT}}}
\newcommand{\lambdabi}{\lambda_{\mathrm{bi}}}
\newcommand{\lambdabox}{\lambda_{\mathrm{box}}}
\newcommand{\Psetn}{\mathcal{P}^{(n)}}

\newcommand{\MDE}{\mathrm{MDE}}
\newcommand{\Mt}{M_{\mathrm{t}}}

\newcommand{\Nset}{\mathcal{N}}
\newcommand{\norm}[1]{\lVert #1 \rVert}
\newcommand{\normFro}[1]{\lVert #1 \rVert_{\mathrm{F}}}

\newcommand{\Pset}{\mathcal{P}}

\newcommand{\RD}{\mathrm{RD}}
\newcommand{\rhat}{\hat{r}}

\newcommand{\SNR}{\mathrm{SNR}}

\newcommand{\Tset}{\mathcal{T}}
\newcommand{\Tc}{T_{\mathrm{c}}}
\newcommand{\Tf}{T_{\mathrm{f}}}
\newcommand{\thetabm}{\bm{\theta}}
\newcommand{\tstamp}[1]{t^{(#1)}}
\newcommand{\TW}{\mathrm{TW}}

\newcommand{\vecrm}[1]{\mathrm{vec}(#1)}

\newcommand{\xbmn}{\bm{\mathrm{x}}^{(n)}}
\newcommand{\Xbmtilde}{\tilde{\bm{\mathrm{X}}}}
\newcommand{\xbmtilde}[1]{\tilde{\bm{\mathrm{x}}}^{(#1)}}
\newcommand{\xbmtildep}{\tilde{\bm{\mathrm{x}}}^{(p)}}

\newcommand{\apc}{a_{\mathrm{c}}}
\newcommand{\apf}{a_{\mathrm{f}}}

\newcommand{\bmp}{\bm{\mathrm{p}}}
\newcommand{\bmpstar}{\bmp^{\star}}

\begin{acronym}[ACRONYM]
\acro{6G}{6th generation wireless systems}
\acro{AoA}{angle-of-arrival}
\acro{AP}{access point}
\acro{APP}{angle-power profile}
\acro{DPP}{delay-power profile}
\acro{AWGN}{additive white Gaussian noise}
\acro{CC}{channel charting}
\acro{CFR}{channel frequency response}
\acro{CSI}{channel-state information}
\acro{CT}{continuity}
\acro{D-MIMO}{distributed multiple input multiple output}
\acro{DFT}{discrete Fourier transform}
\acro{DT}{digital twin}
\acro{GPS}{global positioning system}
\acro{ISAC}{integrated sensing and communications}
\acro{KS}{Kruskal stress}
\acro{LoS}{line-of-sight}
\acro{MDE}{mean distance error}
\acro{MLP}{multi-layer perceptron}
\acro{NLoS}{non-line-of-sight}
\acro{NN}{neural network}
\acro{OFDM}{orthogonal frequency-division multiplexing}
\acro{PDE}{percentile distance error}
\acro{RD}{Rajski distance}
\acro{SL}{supervised learning}
\acro{ToA}{time-of-arrival}
\acro{TDP}{truncated delay profile}
\acro{TW}{trustworthiness}
\acro{UE}{user equipment}
\acro{ULA}{uniform linear array}
\end{acronym}

\begin{document}

\title{Positioning via Digital-Twin-Aided Channel Charting with Large-Scale CSI Features}

\author{
Jos\'{e} Miguel Mateos-Ramos,~\IEEEmembership{Graduate Student Member,~IEEE}, 
Frederik Zumegen,~\IEEEmembership{Graduate Student Member,~IEEE},
Henk Wymeersch,~\IEEEmembership{Fellow,~IEEE},
Christian H\"{a}ger,~\IEEEmembership{Member,~IEEE},
Christoph Studer,~\IEEEmembership{Senior Member,~IEEE},
        % <-this % stops a space
\thanks{%Manuscript received month day, year;revised month day, year. 
The work of JMMR, HW, and CH was supported, in part, by a grant from the Chalmers AI Research Center Consortium (CHAIR), by the National Academic Infrastructure for Supercomputing in Sweden (NAISS), the Swedish Foundation for Strategic Research (SSF) (grant FUS21-0004, SAICOM), and Swedish Research Council (VR grant 2022-03007). The work of CH was also supported by the Swedish Research Council under grant no. 2020-04718. The work of FZ and CS was supported in part by the Swiss National Science Foundation (SNSF) grant 200021\_207314 and by CHIST-ERA grant for the project CHASER (CHIST-ERA-22-WAI-01) through the SNSF grant 20CH21\_218704.}%
\thanks{Jos\'{e} Miguel Mateos-Ramos, Christian H\"{a}ger, and Henk Wymeersch are with the Department of Electrical Engineering, Chalmers University of Technology, Sweden (email: josemi@chalmers.se; christian.haeger@chalmers.se; henkw@chalmers.se).}%
\thanks{Frederik Zumegen and Christoph Studer are with the Department of Information Technology and
Electrical Engineering, ETH Z\"urich, 8092 Z\"urich, Switzerland (email: fzumegen@ethz.ch; cstuder@ethz.ch).}%
}

%\markboth{Journal title}%

% \IEEEpubid{0000--0000/00\$00.00~\copyright~2023 IEEE}

\maketitle
\begin{abstract}
    Channel charting (CC) is a self-supervised positioning technique whose main limitation is that the estimated positions lie in an arbitrary coordinate system that is not aligned with true spatial coordinates.
    In this work, we propose a novel method to produce CC locations in true spatial coordinates with the aid of a digital twin (DT). 
    Our main contribution is a new framework that (i) extracts large-scale channel-state information (CSI) features from estimated CSI and the DT and (ii) matches these features with a cosine-similarity loss function. 
    The DT-aided loss function is then combined with a conventional CC loss to learn a positioning function that provides true spatial coordinates without relying on labeled data. Our results for a simulated indoor scenario demonstrate that the proposed framework reduces the relative mean distance error by 29\% compared to the state of the art. 
    We also show that the proposed approach is robust to DT modeling mismatches and a distribution shift in the testing~data.
\end{abstract}


\section{Introduction} \label{sec:introduction}

Endowing communication networks with positioning capabilities is a key enabler in next-generation wireless systems for applications such as vehicle-to-everything communications, human activity sensing, or smart factories, among others \cite{behravan2023positioning, lu2024integrated}. Conventional positioning methods based on \ac{ToA}, \ac{AoA}, or received signal strength typically assume \ac{LoS} propagation between transmitter and receiver~\cite{liu2007survey, yassin2017recent}. Under \ac{NLoS} conditions, the performance of such methods significantly degrades. 
Alternative solutions that rely on large corpora of ground-truth labeled \ac{UE} positions are robust to \ac{NLoS} scenarios via supervised learning of \acp{NN}~\cite{he2016wifi, wang2017csi}.
The number of labeled positions can be reduced through self-supervised learning, where pseudo-labeled data is first generated to pre-train the \ac{NN}, followed by a fine-tuning phase with few labeled positions~\cite{ott2024radio, salihu2024ssl}. Similarly, semi-supervised learning integrates supervised objectives with additional loss terms that do not require labeled data~\cite{gong2024semisupervised}.
Alternatively, labeled data demands can be reduced via crowdsourcing, where users report position updates using global navigation satellite systems or inertial measurement unit measurements to augment the corpus of labeled samples~\cite{yu2019effective, xiang2021self, tong2021csi, si2025unsupervised}. 
Nevertheless, ground-truth labeled data acquisition may imply a costly measurement campaign, even for a small dataset. Additionally, crowdsourcing positioning relies on the external sensors that often provide inaccurate positioning measurements that can propagate over time~\cite{MURATA201914}. 

To overcome these limitations, \ac{CC} formulates positioning as a self-supervised dimensionality-reduction problem, mapping high-dimensional \ac{CSI} to lower-dimensional position estimates \cite{studer2018channel,ferrand2023wireless}. The key principle of CC is to preserve local geometry, ensuring that nearby estimated pseudo-positions correspond to nearby true positions. However, this approach yields estimated pseudo-positions in an arbitrary coordinate system, i.e., global geometry may not be preserved. Hence, there is a need to develop CC-based localization methods that provide coordinates in a global reference system, i.e., in true spatial coordinates. 

As one possible approach to obtain coordinates in a global reference system, some prior works incorporate a small degree of ground-truth UE positions into the CC training objective. 
This can be done, for example, by adding the distance error between the pseudo-positions and the true positions to standard CC loss functions that preserve local geometry~\cite{huang2019improving, lei2019siamese, zhang2021semi, deng2021network, viet2022implicit, palhares2025csi2vec}. 
Another approach is to use the ground-truth positions as anchor points to compute an optimal affine transformation, with the goal of aligning the estimated positions with the real coordinate system~\cite{stahlke2023indoor, euchner2024uncertainty, zhao2024signature, stephan2024angle}. 
However, 
labeled data collection still implies a measurement campaign.
Moreover, new labeled data needs to be collected for every different environment.
Another approach is to combine knowledge of access point (AP) locations with knowledge of LoS zones and path-loss models to embed pseudo-positions in real-world coordinates~\cite{taner2025channel}.
In contrast to such methods, we will focus on CC-based positioning methods that deliver true spatial coordinates without requiring any ground-truth position labels.

\subsection{Relevant Prior Art} \label{subsec:prior_art}

Existing ``label-free'' CC approaches that provide coordinates in a global reference system can be divided into the following five categories: (i) based on sensor fusion~\cite{esrafilian2024global}, (ii) based on \acp{DT}~\cite{cazzella2025chartwin}, (iii) based on affine transformations leveraging floor plans~\cite{pihlajasalo2020absolute, stahlke2024velocity, zhang2025uniloc, vindas2024multi}, (iv)  based on conventional estimation methods~\cite{euchner2023augmenting, euchner2024leveraging}, and (v) based on path-loss models \cite{karmanov2021wicluster, taner2025channel}.

The work in \cite{esrafilian2024global}  introduces a sensor-fusion approach where the UEs are equipped with a two-dimensional laser scanner that collects depth measurements. The displacement of the UEs between time steps is estimated from the laser data and added to a tailored loss function during training to preserve global geometry.
However, including a laser scanner in the UEs is costly and positioning requires tight synchronization between the UEs and the APs, which is impractical. 

In \cite{cazzella2025chartwin}, a digital representation of the physical environment, or \textit{\ac{DT}}~\cite{tao2019digital}, is used to preserve global geometry, avoiding the need for external sensors in the UEs. An outdoor urban environment is simulated with a DT to generate synthetic labeled positioning data that relates the CSI in the DT to ground-truth positions. DT simulations include detailed building and traffic information to obtain accurate synthetic data that is less costly to collect than real measurements. 
However, the fine-grained details in the DT drastically increase the modeling complexity of the scenario, the computational complexity of the simulations, and the likelihood of degrading positioning performance under model mismatches.

Other works leverage affine transformations without ground-truth positions to align the estimated positions with the real coordinate system \cite{pihlajasalo2020absolute, stahlke2024velocity, zhang2025uniloc, vindas2024multi}. 
These approaches typically do not require fine-grained details of the environment and are easier to implement in different scenarios.
In~\cite{pihlajasalo2020absolute}, AP positions are estimated based on the measurements with the highest received power, which are then used as anchors to find the optimal transform.
The method in~\cite{stahlke2024velocity} extends the optimal transport in~\cite{ghazvinian2021modality}, which aligns the estimated positions with the floor plan, to also learn the probabilities that positions fall within specific regions. This approach enables uncertainty quantification of the estimated position, e.g., for estimated positions belonging to unexplored areas.
In~\cite{zhang2025uniloc}, model-based and \ac{NN}-based methods are combined, leveraging model-based estimates under LoS conditions and training a NN in a self-supervised manner to estimate NLoS samples. The self-generated labeled data for the NN consists of model-based estimations, where an optimal transport operation is also performed in the case of NLoS samples.
In~\cite{vindas2024multi}, the coverage area of each AP is known and the optimal transport operation maps the pseudo-positions estimated by each AP to their coverage area while preserving the overlap between the coverage areas of the APs.
The first limitation of \cite{pihlajasalo2020absolute, stahlke2024velocity, zhang2025uniloc, vindas2024multi} is that they require a large number of estimated positions to apply the affine transform and obtain accurate performance. The approach in \cite{pihlajasalo2020absolute} needs to distinguish samples close to the APs, \cite{stahlke2024velocity} constructs a probability map of the estimated positions, and \cite{zhang2025uniloc} assumes that NLoS estimates follow a uniform distribution across the NLoS area. Under a limited number of estimated positions, e.g., if the UEs have not explored the entire environment, the performance of such methods likely degrades. Additionally, the considered transforms only apply affine operations, disregarding more complex non-affine mappings.

The work in~\cite{euchner2023augmenting} designs a loss function based on the root-multiple signal classification algorithm for \ac{AoA} and \ac{ToA} estimation. This loss function is added to a CC loss that preserves the local geometry. In~\cite{euchner2024leveraging}, the Doppler effect is included in a tailored loss function assuming a free-space channel model and knowledge of the AP positions.
The estimated positions of~\cite{euchner2023augmenting, euchner2024leveraging} in global coordinates are directly obtained in real time without relying on affine transformations and  conventional algorithms are outperformed, but the performance under NLoS conditions remains unclear.

Finally, works based on path-loss models~\cite{karmanov2021wicluster, taner2025channel} assume that AP positions are known and enforce estimated positions to lie closer to the AP with the strongest received signal. 
For instance,~\cite{karmanov2021wicluster} divides the floor plan into zones and leverages a small subset of zone-annotated CSI samples during training to predict the zone of new samples. Once a new sample is assigned to a zone, two weakly supervised losses enforce it to belong to the estimated zone and lie closer to \acp{AP} with the strongest received power. 
However, the method in~\cite{karmanov2021wicluster} requires zone-annotated data, which increases the complexity of the positioning algorithm and zone-division is based on a heuristic method.
The recent work of \cite{taner2025channel} performs global positioning without dividing the floor plan into zones or annotated data. Instead, a power threshold based on received signals is computed to distinguish measurements in the LoS area of each AP and estimated positions are, using a bilateration loss, enforced to be closer to the AP that received the strongest signal. Additionally, if a bounding box describing the LoS area of each AP is available, LoS positions are further constrained to remain within these bounding boxes.
Although the work in~\cite{taner2025channel} provides an effective way of estimating positions in global coordinates, it presents several issues that limit its practical implementation: 
Firstly, distinguishing if a received signal belongs to the LoS area of an AP is based on thresholding the received power at each AP, which can fail in scenarios 
with heterogeneous devices that can select their transmission power. Secondly, the bilateration loss is based on path-loss models that fail under NLoS conditions. Therefore, there is a need to develop self-supervised positioning techniques that can work with different devices and under NLoS propagation conditions.

\subsection{Contributions} \label{subsec:contributions}
In this work, we leverage a DT of an uplink \ac{D-MIMO} scenario with several \acp{UE} and \acp{AP} in order to perform CC that preserves the global geometry of estimated positions.
The key contributions of this work are as follows:
\begin{itemize}
    \item \emph{Efficient DT-aided self-supervised positioning:} We propose, for the first time, a computationally efficient self-supervised DT-aided positioning framework that can work with heterogeneous devices. Our method is based on large-scale features computed from the DT and the CSI. Recent work has shown that large-scale features can be computed in a DT for real-time applications \cite{ait2024fast}. 
    As large-scale features, we consider the average received power, the power per angular direction and delay tap, the truncated delay profile, and the magnitude of the covariance matrix of the channel. 
    Compared to \cite{taner2025channel}, which is based on thresholding the power of the received signals at each AP, our large-scale features do not depend on the transmitted power. This advantage enables our method to work with heterogeneous devices that freely select their transmit powers.
    Moreover, unlike~\cite{taner2025channel}, our method constrains the estimated positions, providing a bounded positioning error.
    Our results show that our approach outperforms the work in \cite{taner2025channel} for the same simulated scenario.
    At the same time, our proposed framework is based on large-scale features that drastically reduce the DT modeling and computational complexity compared to \cite{cazzella2025chartwin}.         
    \item \emph{Robustness under modeling mismatches: } We evaluate the performance of our approach under mismatches in the modeling of the DT compared to the real scenario, in contrast to \cite{cazzella2025chartwin}, in which the real and simulated data are drawn from the same DT simulation. 
    We consider mismatches in the modeling of the position of the APs during training. We show that even under such mismatches, the proposed framework remains robust and obtains better performance compared to the state of the art. 
    \item \emph{Generalization capability:} We consider a distribution shift between the training and the testing data. Although the proposed framework is continuously trained, this test evaluates the performance when the scenario changes before the system is fine-tuned with new observed data, which is rarely studied in CC works. 
    In our simulations, we decrease the height of the UEs during inference and show that the proposed approach only slightly degrades under such mismatches when large-scale features are considered.
\end{itemize}

\subsection{Outline}
The rest of the paper is organized as follows. Sec.~\ref{sec:sys_model} introduces the system model, including the DT description. Sec.~\ref{sec:dt_aided_cc} presents a background on CC and proposes our framework. Sec.~\ref{sec:results} compares the proposed approach with state-of-the-art baselines and Sec.~\ref{sec:conclusions} concludes. 

\subsection{Notation}

Vectors and matrices are denoted as boldface lowercase and uppercase letters, respectively. Sets are denoted as uppercase calligraphic letters and are defined by curly brackets; the cardinality of a set $\mathcal{A}$ is denoted as $|\mathcal{A}|$. The transpose and conjugate transpose of a matrix $\Abm$ are denoted by $\Abm^{\top}$ and~$\Abm^{\hermit}$, respectively. The $i$th row and $j$th column of a matrix $\Abm$ are denoted as $[\Abm]_{i,:}$ and $[\Abm]_{:,j}$, respectively. The upper-left submatrix containing the first $M$ rows and $N$ columns of $\Abm$ is denoted as $[\Abm]_{1:M, 1:N}$. The matrix vectorization operation is denoted as $\vecrm{\Abm}$. We denote the aggregation of two matrices $\Abm\in\realset[M][N]$ and $\Bbm\in\realset[M][N]$ along the first dimension as $[\Abm;\Bbm]\in\realset[2M][N]$ and along the second dimension as $[\Abm,\Bbm]\in\realset[M][2N]$. The positive part of a real number $x$ is denoted as $(x)^+=\max\{0,x\}$. The $N$-point unitary \ac{DFT} matrix is denoted as $\dft{N}$. The $\ell^2$-norm of a vector and the Frobenius norm of a matrix are denoted as $\norm{\abm}$ and $\normFro{\Abm}$, respectively; the all-ones vector is denoted as~$\bm{1}$ and the indicator function is denoted as $\mathds{1}\{\cdot\}$.

\section{System Model} \label{sec:sys_model}
We consider an uplink \ac{D-MIMO} scenario with multiple \acp{UE} and $A$ \acp{AP}. The \acp{UE} transmit \ac{OFDM} signals with $S$ subcarriers at timestamp $\tstamp{n}$ from the position $\xbm[n]$, for $n\in\Nset=\{1, \ldots, N\}$. Each \ac{AP} and \ac{UE} is equipped with an antenna array of $K$ and $L$ antenna elements, respectively. 
We assume a multiple-access protocol that schedules the UE transmissions, ensuring that the received signal at each AP and timestamp corresponds to a single UE.
The \ac{CSI} matrix estimated by AP $a$ at the $n$th timestamp $\tstamp{n}$ is $\Hbm[a][n]\in\complexset[K][S]$, for $a=1,2,\ldots,A$.
The estimated CSI matrix includes the effect of \ac{AWGN} of variance $\sigma^2$. The \ac{CSI} matrix at timestamp $\tstamp{n}$ is $\Hbm[n] = [\Hbm[1][n];\cdots;\Hbm[A][n]]~\in~\complexset[M][S]$, with $M=AK$. The collection $\{\Hbm[n], \tstamp{n}\}_{n=1}^N$ constitutes the database of data collected from the real environment. 

We also consider a \ac{DT} that models the uplink \ac{D-MIMO} scenario. 
The DT is defined by the position and materials of reflective surfaces (walls, floor, ceiling, or other objects), the position of the \acp{AP}, and radio-frequency parameters (carrier frequency, bandwidth, antennas, and transmitted signal). 
We predefine a set of UE positions in the \ac{DT} $\{\xbmtildep\}_{p=1}^P$, for $p=1,\ldots,P$, where $\xbmtildep\in\realset[\Dx]$ and $\Dx$ is either 2 or 3.  
The CSI matrix synthesized from the \ac{DT} at AP $a$ corresponding to the $p$th position is denoted by $\Hbmtilde[a,p]\in\complexset[K][S]$ and the CSI matrix corresponding to the $p$th position is $\Hbmtilde[p]=[\Hbmtilde[1,p];\ldots;\Hbmtilde[A,p]] \in\complexset[M][S]$. The collection $\{\Hbmtilde[p], \xbmtilde{p}\}_{p=1}^P$ constitutes the database of simulated CSI and predefined positions from the DT. 
As mentioned in Sec.~\ref{subsec:contributions}, our proposed framework is based on large-scale features from the DT and the observed CSI matrix.
Generally, DTs can simulate the CSI matrix observed by each AP and we can then compute large-scale features from the CSI matrix, but efficient DTs can directly provide large-scale features; see, e.g.,~\cite{ait2024fast}.
We based the description of the DT and the proposed method of Sec.~\ref{sec:dt_aided_cc} on the general case that the DT simulates the CSI matrix, but in Sec.~\ref{sec:results} we also consider large-scale features directly computed from the DT.

The objective of positioning is to find a function that maps the estimated CSI from the measured data  $\Hbm[n]$ to the true UE position $\xbm[n]$, given some side information about the environment through the DT model. To find such a function, we first design a training phase that uses the unlabeled dataset $\{\Hbm[n], \tstamp{n}\}_{n=1}^N$ and the synthesized DT data $\{\Hbmtilde[p], \xbmtilde{p}\}_{p=1}^P$ to produce estimated positions close to the true positions.

\section{Digital-Twin-Aided Channel Charting} \label{sec:dt_aided_cc}

We now describe the operating principles of conventional CC, which serves as a foundation for our proposed method. We then describe the proposed framework and formulate the training procedure as an optimization problem. We conclude with a detailed description of the considered large-scale features.

\subsection{Background on Channel Charting} \label{subsec:background_cc}
\Ac{CC} is composed of two operations~\cite{studer2018channel}: (i) feature extraction and (ii) positioning \ac{CC} function. In (i), large-scale feature vectors are computed from the \ac{CSI} since small-scale fading can produce large variations in the \ac{CSI} for close positions in space~\cite{goldsmith2005wireless}. We compute the large-scale CSI feature vector $\fbm[n]$ at time instant~$\tstamp{n}$ as in~\cite{taner2025channel} as follows:
\begin{align} \label{eq:input_nn}
    \fbm[n] = f(\Hbm[n]) = \frac{|\hbm[n]|}{\norm{\hbm[n]}}.
\end{align}
Here, $\hbm[n] = \vecrm{[\Hbm[n] \dft{S}^{\hermit}]_{:,1:C}}\in\complexset[\Dh]$ and $\Dh=MC$.
An inverse \ac{DFT} of the \ac{CSI} matrix across subcarriers is performed in~\cite{taner2025channel} because most of the received power is expected to be on the first $C$ time-domain taps and the dimensionality of~$\Hbm[n]$ can be reduced with minimal information loss.
In (ii), the positioning \ac{CC} function is expressed as $\xbmhat[n]=\gtheta{\fbm[n]}\in\realset[\Dx]$ with learnable parameters $\thetabm$. This function can then be used to extract the estimated positions of the \acp{UE}.

The goal of \ac{CC} is to optimize the parameters $\thetabm$ so that training samples with similar feature vectors $\fbm[n]$ yield estimated positions $\xbmhat[n]$ that are close in $\realset[\Dx]$. Different loss functions have been proposed to optimize the parameters $\thetabm$~\cite{studer2018channel, ferrand2021triplet, huang2019improving, lei2019siamese, rappaport2021improving, yassine2022leveraging}. In this work, we utilize the timestamp-based triplet-loss approach from \cite{ferrand2021triplet} to enable a fair comparison with \cite{taner2025channel} (see Sec.~\ref{sec:results} for the details), which we describe in the following. First, we define the set of triplets
\begin{align} \label{eq:T_set}
    \Tset = \{(&n,c,f)\in\Nset^3: \nonumber \\
    &0 < |\tstamp{n}-\tstamp{c}|\leq \Tc \leq \Tf < |\tstamp{n}-\tstamp{f}|\},
\end{align}
where $\Tc,\Tf >0$ are coherence-time parameters that distinguish if \ac{CSI} measurements are close ($\Tc$) or far ($\Tf$) in time to a CSI measurement at a timestamp $\tstamp{n}$. If $|\tstamp{n}-\tstamp{c}| < |\tstamp{n}-\tstamp{f}|$, we expect that $\xbmhat[n]$ is closer to $\xbmhat[c]$ than to $\xbmhat[f]$. 
Second, the timestamp-based triplet loss is defined as
\begin{align} \label{eq:triplet_loss}
    \Lfrak[CC](\thetabm) = \frac{1}{|\Tset|} \sum_{(n,c,f)\in\Tset} &(\norm{\xbmhat[n]-\xbmhat[c]} \nonumber \\
    &- \norm{\xbmhat[n]-\xbmhat[f]}+\Mt)^+,
\end{align}
where $\Mt\geq 0$ is a hyper-parameter that enforces $\xbmhat[n]$ to be at least $\Mt$ closer to $\xbmhat[c]$ than to $\xbmhat[f]$. Conventional CC aims to solve the optimization problem
\begin{align} \label{eq:triplet_problem}
    \thetabm^{\star} = \arg\min_{\thetabm} \Lfrak[CC](\thetabm),
\end{align}
where $\thetabm^\star$ are the optimal parameters of the positioning CC function~$\gtheta{\cdot}$.
The problem in~\eqref{eq:triplet_problem} aims to preserve local geometry, but there is no guarantee that the learned CC positioning function reduces the error between the true and estimated positions. To resolve this limitation, we now introduce \ac{DT}-aided CC .

\subsection{Proposed Method} \label{subsec:proposed_method}
\begin{figure*}[tb]
    \centering
    \input{figures/block_diagram}
    \vspace{0.2cm}
    \caption{Block diagram of the proposed approach. Dashed elements are predefined and kept fixed during training. The large-scale features in the DT can be computed directly without first obtaining the CSI $\Hbmtilde[p]$. During inference, the DT and the DT-aided learning blocks are not used.}
    \label{fig:proposed_approach}
\end{figure*}
The proposed method is based on conventional CC and the timestamp-based triplet-loss of \eqref{eq:triplet_loss}. We include the DT in CC in two stages: (i) in conventional CC to confine the estimated positions to the convex hull of the predefined DT positions and (ii) adding a new loss function to the loss of \eqref{eq:triplet_loss} that aims to preserve  global geometry. 
The proposed framework is represented in Fig.~\ref{fig:proposed_approach} and described in the following.

\subsubsection{Modified Channel Charting}
Firstly, we turn the conventional CC regression problem into a classification problem as shown in the green box of Fig.~\ref{fig:proposed_approach}. Instead of estimating the positions as $\xbmhat[n]=\gtheta{\fbm[n]}$, we increase the dimensionality of the output of the learnable function as in~\cite{gonultas2022csi} and estimate the probability mass function $\pbm[n]=\gtheta{\fbm[n]} \in[0,1]^P$ of the true position at $\tstamp{n}$, i.e., $[\pbm[n]]_i$ is the probability that the $i$th predefined position $\xbmtilde{i}$ is the true position at $\tstamp{n}$ and $\bm{1}^{\top}\pbm[n]=1$.
Treating the localization problem as a classification problem was shown in~\cite{gonultas2022csi} to result in better positioning performance with supervised learning. 
We then compute the expected position as $\xbmhat[n] = q(\pbm[n]) = \Xbmtilde\pbm[n]$, with $\Xbmtilde = [\xbmtilde{1}, \ldots, \xbmtilde{P}]\in\realset[\Dx][P]$. 
This mapping allows us to use the triplet loss in \eqref{eq:triplet_loss} to preserve local geometry and the estimated position is confined to the convex hull of the predefined positions $\Xbmtilde$. 

\subsubsection{Digital-Twin-Aided Loss Function}
Secondly, we introduce a novel DT-aided loss function to preserve global geometry, corresponding to the orange box in Fig.~\ref{fig:proposed_approach}. 
The proposed loss function is based on large-scale features computed from the observed CSI and the DT. Large-scale features are expected to vary only slowly with the position of the UEs, while the CSI rapidly varies over space in the order of the wavelength. 
Moreover, learnable functions like the \ac{NN} considered in this work (see Sec.~\ref{subsection:scenario}) suffer from spectral bias for lower frequencies \cite{rahaman2019spectral, cao2021towards, zhiqin2020frequency}, which correspond to slowly varying features in our setup.
From the DT-based CSI $\Hbmtilde[a,p]$, the feature vector estimated by \ac{AP} $a$ corresponding to the $p$th position in the \ac{DT} is $\vbmbar[a,p]=h(\Hbmtilde[a,p])\in\complexset[D]$, for $p=1,2,\ldots,P$, where $D$ is the dimension of the feature vector to consider (see Sec.~\ref{subsec:features}). 
We denote the function that computes the large-scale features generally as $h(\cdot)$ since it can be different from $f(\cdot)$.
The feature vector for the $p$-th predefined position is $\vbmbar[p]=[\vbmbar[1,p];\cdots;\vbmbar[A,p]]\in\complexset[\Dv]$, where $\Dv=AD$.
We compute the expected value of the large-scale feature vectors $\{\vbmbar[p]\}_{p=1}^P$ at timestamp $\tstamp{n}$ using the estimated probabilities $\pbm[n]$ as $\vbmtilde[n] = u(\pbm[n]) = \Vbmbar\pbm[n]\in\complexset[\Dv]$, where $\Vbmbar = [\vbmbar[1], \ldots, \vbmbar[P]]\in\complexset[\Dv][P]$. 
Similarly, from the measured CSI we compute a large-scale feature vector per \ac{AP} at $\tstamp{n}$ as $\vbm[a,n] = h(\Hbm[a,n])\in\complexset[D]$. Aggregating across \acp{AP}, we obtain another feature vector at $\tstamp{n}$ as $\vbm[n] = [\vbm[1,n]; \ldots; \vbm[A,n]]\in\complexset[\Dv]$. 
We combine the feature vectors from the \ac{DT} and the measured \ac{CSI} in the loss function
\begin{align} \label{eq:DT_loss}
    \Lfrak[DT](\thetabm) = \frac{1}{N} \sum_{n\in\Nset} \exp\bigg( - \frac{|(\vbm[n])^{\hermit}\vbmtilde[n]|^2}{\norm{\vbm[n]}^2 \norm{\vbmtilde[n]}^2} \bigg).
\end{align}
The loss function in \eqref{eq:DT_loss} involves the cosine similarity between the vectors~$\vbm[n]$ and $\vbmtilde[n]$, which is used in other deep learning tasks~\cite{Barz_2020_WACV, NEURIPS2021_cbcb58ac}, and a nonlinear exponential term that empirically provides better results in our scenario.
Moreover, the vector normalization in \eqref{eq:DT_loss} accounts for mismatched real and simulated transmitted powers of the UEs.
Note that the feature vector $\vbm[n]$ from the estimated \ac{CSI} does not depend on the learnable parameters as the function $h(\cdot)$ is not learnable. 
The loss in \eqref{eq:DT_loss} provides a theoretical guarantee of the positioning performance when the synthesized DT CSI~$\Hbmtilde[p]$ matches the estimated CSI $\Hbm[n]$.

\begin{lemma} \label{lemma:min_DT_loss}
    The loss function in \eqref{eq:DT_loss} is minimized if and only if~$\vbm[n]$ is linearly dependent with $\vbmtilde[n]$.
\end{lemma}

\begin{proof}
    \begin{align}
        \Lfrak[DT](\thetabm) &= \frac{1}{N} \sum_{n\in\Nset} \exp\bigg( - \frac{|(\vbm[n])^{\hermit}\vbmtilde[n]|^2}{\norm{\vbm[n]}^2 \norm{\vbmtilde[n]}^2} \bigg) \\
        &\geq \exp\bigg( - \frac{1}{N} \sum_{n\in\Nset} \frac{|(\vbm[n])^{\hermit}\vbmtilde[n]|^2}{\norm{\vbm[n]}^2 \norm{\vbmtilde[n]}^2}  \bigg) \label{proof:lemma_jensen} \\
        &\geq \exp(-1), \label{proof:cs}
    \end{align}
    where in \eqref{proof:lemma_jensen} we used Jensen's inequality~\cite{kuczma2009introduction} and in \eqref{proof:cs} we used Cauchy--Schwarz inequality~\cite{laub2004matrix}. The inequalities in \eqref{proof:lemma_jensen} and~\eqref{proof:cs} hold with equality if and only if $\vbm[n]$ is linearly dependent with $\vbmtilde[n]$.
\end{proof}

\begin{proposition} \label{prop:exact_est}
    Let 
    \begin{align}
        \Lset[P] = \{\bmp \in [0,1]^P: \bm{1}^{\top}\bmp=1\}
    \end{align}
    be the set of all possible probability vectors of dimension $P$. Assume a matched DT such that 
    \begin{align}
        \xbm[n] &= \Xbmtilde \bmpstar \label{eq:prop_1_cond_x} \\
        \vbm[n] &= \Vbmbar\bmpstar \neq 0, \label{eq:prop_1_cond_v}
    \end{align}
    for some $\bmpstar\in\Lset[P]$ and a fixed $n\in\Nset$. Suppose $\Vbmbar\in\complexset[\Dv][P]$ is tall and has full rank ($P\leq\Dv$). Then, any global minimizer~$\thetabm^{\star}$ satisfies
    \begin{align}
        \pbmstar[n] = \bmpstar \quad \text{and} \quad \xbmhatstar[n] = \Xbmtilde \pbmstar[n] = \xbm[n].
    \end{align}
\end{proposition}
\begin{proof}
    By Lemma~\ref{lemma:min_DT_loss}, at any global minimizer $\thetabm^{\star}$ we must have
    \begin{align}
        \vbmtildestar[n] = \Vbmbar\pbmstar[n] = \alpha\vbm[n]=\alpha\Vbmbar \bmpstar,
    \end{align}
    for some $\alpha\in\complexset\backslash\{0\}$. Because $\Vbmbar$ is full rank, the mapping $\bmp \mapsto \Vbmbar\bmp$ is injective, implying that
    \begin{align}
        \pbmstar[n] = \alpha\bmpstar.
    \end{align}
    Since both $\pbmstar[n]$ and $\bmpstar$ belong to $\Lset[P]$, it follows that $\alpha=1$ and hence, $\pbmstar[n]=\bmpstar$. Substituting into $\xbmhatstar[n] = \Xbmtilde\pbmstar[n]$ yields $\xbmhatstar[n] = \xbm[n]$.
\end{proof}
\begin{remark}
    Because $\Xbmtilde$ is typically a wide matrix~($P>\Dx$), the mapping $\bmp\mapsto \Xbmtilde\bmp$ is not injective---in fact, many $\bmp\in\Lset[P]$ can represent the same position $\xbm[n]=\Xbmtilde \bmp$.
    Hence, barycentric coordinates in position space are generally non-unique. However, since $\Vbmbar$ is tall and full column rank, the mapping $\bmp\mapsto\Vbmbar\bmp$ is injective. Consequently, even if several~$\bmp$ vectors lead to the same $\xbm[n]$, only one of them yields the correct feature vector $\vbm[n]=\Vbmbar \bmpstar$. The optimization in feature space therefore selects the unique $\bmpstar$ consistent with the observed $\vbm[n]$, leading to $\xbmhatstar[n] = \Xbmtilde \bmpstar=\xbm[n]$. In this sense, the tall matrix $\Vbmbar$ disambiguates the non-uniqueness introduced by the wide matrix~$\Xbmtilde$.
\end{remark}

Although Proposition~\ref{prop:exact_est} guarantees that a global optimizer that minimizes \eqref{eq:DT_loss} can yield a perfect estimation under a matched DT, fulfilling condition \eqref{eq:prop_1_cond_v} for all $n\in\Nset$ while $\Vbmbar$ is full rank is not straightforward, as it depends on the environment, the AP positions, and the large-scale features to consider. Given that our goal is to propose a general framework that can work for any given environment, we propose to combine  
the loss functions of \eqref{eq:triplet_loss} and \eqref{eq:DT_loss} as 
\begin{align} \label{eq:total_loss}
    \Lfrak(\thetabm) = \lambdaCC\Lfrak[CC](\thetabm) + \lambdadt \Lfrak[DT](\thetabm),
\end{align}
where $\lambdaCC, \lambdadt\geq0$ are tunable weights. Based on $\Lfrak(\thetabm)$, we aim to solve 
\begin{align} \label{eq:proposed_problem}
    \thetabm^\star = \arg\min_{\thetabm} \Lfrak(\thetabm)
\end{align}
to learn a positioning function~$\gtheta{\cdot}$ that maps CSI features to positions close to the true positions. The proposed approach is summarized in Algorithm~\ref{alg:training} for a fixed data set. If new samples of \ac{CSI} are estimated, the same algorithm can be continuously applied with an extended data set including the new data points.

\begin{algorithm}[tb]
	\caption{Training algorithm of the proposed approach.}
	\label{alg:training}
    \begin{algorithmic}[1]
	    \State \textbf{Input:} Data set of estimated CSI $\{\Hbm[n]\}_{n=1}^N$, grid of points in the \ac{DT} $\Xbmtilde$, large-scale features from the \ac{DT} $\Vbmbar$.
        \State \textbf{Output:} Optimized \ac{NN} parameters $\thetabm^{(N_{\mathrm{iter}})}$.
        \State \textbf{Initialization:} \ac{NN} Parameters $\thetabm^{(0)}$.
        \For{$i=0,\ \ldots,\ N_{\mathrm{iter}}-1$}
            \State Compute large-scale feature  $\vbm[n] = h(\Hbm[n])$.
            \State Compute the CSI features $\fbm[n]$ following~\eqref{eq:input_nn}.
            \State Estimation with \ac{NN}: $\pbmi[n]=g_{\thetabm^{(i)}}(\fbm[n])$.
            \State Compute $\xbmhati = \Xbmtilde \pbmi[n]$.
            \State Average large-scale features $\vbmtildei[n] = \Vbmbar \pbmi[n]$.
            \State Calculate $\Lfrak[CC](\thetabm^{(i)})$ following~\eqref{eq:triplet_loss}.
            \State Compute \ac{DT} loss $\Lfrak[DT](\thetabm^{(i)})$ in~\eqref{eq:DT_loss}.
            \State Compute $\Lfrak(\thetabm^{(i)})$ according to~\eqref{eq:total_loss}.
            \State Update $\thetabm^{(i)}$ based on gradient descent.
        \EndFor
	\end{algorithmic}
\end{algorithm}

\subsection{Large-Scale Features} \label{subsec:features}
Here we propose different large-scale feature processing functions $h(\cdot)$. These functions determine the large-scale features to compute from the estimated CSI and by the DT.

\subsubsection{Power} \label{subsubsec:power}
We consider that the power received by AP $a$ at timestamp $\tstamp{n}$ is $h(\Hbm[a,n]) = \norm{\hbm[a,n]}^2$, where $\hbm[a,n]=\vecrm{[\Hbm[a,n] \dft{S}^{\hermit}]_{:,1:C}}$, which means that $D=1$. This is the most straightforward large-scale feature vector and the easiest to acquire. The advantage is that this feature vector also works in a simpler scenario where the \acp{AP} have a single antenna element. 
However, the \textit{Power} feature presents a disadvantage as proven in the following proposition.
\begin{proposition} \label{prop:power}
    Let 
    \begin{align}
        \Pset^{(a)} = \{p\in\{1,\ldots,P\}: \norm{\Hbmtilde[a',p]}_F^2=0 \text{ for } a'\neq a\}
    \end{align}
    be the set of points in the DT whose transmitted signal is at most received by AP $a$.
    If $[\pbm[n]]_p=0$ for $p\notin\Pset^{(a)}$, then
    $\Lfrak[DT]$ does not depend on $\thetabm$.
\end{proposition}
\begin{proof}
    The claim follows from the fact that for points $p\in\Pset^{(a)}$, $\vbmbar[p]$ is nonzero in the $a$th position, and the normalized vector $\vbmtilde[n]/\norm{\vbmtilde[n]}$ in \eqref{eq:DT_loss} does not depend on $\thetabm$.
\end{proof}
Proposition~\ref{prop:power} implies that if there is a subset of points in the DT that are only served by one AP, the learnable function $\gtheta{\cdot}$ stops updating the parameters $\thetabm$ when the estimated position~$\xbmtilde{n}$ is a combination of that subset of points. In areas covered by just one AP, the positioning error is therefore likely to be large.
This means that the signal from the \acp{UE} must be received in the DT by at least two \acp{AP} to reduce the positioning error.\footnote{Note that this does not mean that the \acp{UE} must be in \ac{LoS} with the \acp{AP}, but the signal transmitted by the UEs must be received by at least two APs.}

\subsubsection{\Ac{APP}}
In order to attempt to resolve the issue of the \textit{Power} feature, we define the \textit{APP} of AP $a$ at $\tstamp{n}$ as $h(\Hbm[a,n])= |\Hbm[a,n]_{\mathrm{F}}|^2\bm{1}$, with $\Hbm[a,n]_{\mathrm{F}} = \dft{K}\Hbm[a,n]$, which means that $D=K$. The APP aims to compute the power of the \ac{CSI} matrix for different angles. This approach is much less likely to suffer the same problem as the \textit{Power} feature if $K>1$ since the dimensionality of $\vbm[n]$ and $\vbmtilde[n]$ increases. In the case of single-antenna APs, the APP is equivalent to the \textit{Power} feature.

\subsubsection{\Ac{DPP}}
Similarly to the \textit{APP}, we define the \textit{DPP} of AP $a$ at $\tstamp{n}$ as $h(\Hbm[a,n])= \bm{1}^{\top}|\Hbm[a,n]\dft{S}^{\hermit}|^2$, which aims to compute the power of the CSI matrix for different delay taps. 
Note that here we consider all $S$ delay taps compared to the $C$ taps in \eqref{eq:input_nn} because the large-scale features are only used to compute the loss in \eqref{eq:DT_loss}, whereas the CSI features $\fbm[n]$ are processed by the positioning function $\gtheta{\cdot}$, which is more computationally demanding.
We include this feature to compare the effect of using angular and delay information in the proposed framework. In this case $D=S$. Assuming $S>K$ (which is typically the case in practice), this feature increases the dimensionality of $\vbm[n]$ compared to the \textit{APP}. 
For single-antenna APs, the \textit{DPP} amounts to the \textit{Power} feature.

\subsubsection{Covariance}
We define the covariance vector of AP $a$ at $\tstamp{n}$ as $h(\Hbm[a,n])=\vecrm{|\Hbm[a,n]_{\mathrm{F}}(\Hbm[a,n]_{\mathrm{F}})^{\hermit}|}$, which means that $D=K^2$. We introduce this feature since second-order moments have previously been used for conventional CC~\cite{studer2018channel}. Note that the \textit{APP} values are contained in $\vbm[n]$. Assuming~\mbox{$K>1$}, this feature increases the dimensionality of $\vbm[n]$ compared to the \textit{APP}.

\subsubsection{\Ac{TDP}}
We follow \eqref{eq:input_nn} for the CSI of each AP, i.e., $h(\Hbm[a,n])=|\hbm[a,n]|/\norm{\hbm[a,n]}$, where $\hbm[a,n] = \vecrm{[\Hbm[a,n]\dft{S}^{\hermit}]_{:,1:C}}$. This considers the large-scale feature of~\cite{taner2025channel} both as input to the positioning function and for the DT-aided loss function. In this case, $D=KC$.

\subsection{Discussion on Digital Twin Implementation} \label{subsec:DT_implementation}
Our proposed approach relies on a DT to perform positioning. The DT comprises three components: the real space, the virtual space, and a communication link between them~\cite{tao2019digital, JONES202036}.
The virtual space consists of a three-dimensional representation of the real environment to acquire simulated data, which in turn includes geometrical and material information. Geometrical modeling of the environment can be constructed via manual measurements, camera information (see~\cite{li20223Dreconstruction} and references therein), or laser scanners~\cite{keqiang2009automated}. Material properties of objects can be estimated through additional measurements~\cite{scott2014material, majumdarmultiphysics}, however, they imply extra costs and may not scale well to large environments. We instead consider that materials are available a priori (e.g., from building specifications) and minor mismatches in material properties can be mitigated by our proposed approach, as shown in Sec.~\ref{sec:results}. After its initial construction, the DT is updated based on information collected from external sensors~\cite{bae2025DT}. The communication link between the DT and the real environment ensures that the DT, and consequently the large-scale features, are updated and that the predicted positions of our proposed approach are transmitted to the APs.

\section{Results} \label{sec:results}

We now present the results to assess the positioning performance of the proposed \ac{DT}-aided \ac{CC} approach.\footnote{Source code to reproduce the simulation results in this paper is available at \url{https://github.com/josemateosramos/DT_CC}.} We first describe the state-of-the-art baselines and the considered evaluation metrics. We then assess the achieved positioning results, the effect of the density of points in the DT, and the robustness under modeling mismatches in the DT. 

\subsection{Baselines} \label{subsec:baselines}

We consider three baselines to compare with our proposed approach. The first baseline is the best-performing CSI-based positioning method that does not rely on labeled data, synchronized APs, or affine transformations. The goal of this baseline is to frame the proposed approach with respect to state-of-the-art CC. The second baseline is a semi-supervised approach that uses labeled data after a first CC learning phase to compute the optimal affine transformation that obtains global coordinates. The last baseline is a supervised CSI fingerprinting approach that serves as a lower bound on the positioning error. Although the last two baselines use additional information in terms of labeled data compared to the proposed approach, we include them to assess how close the proposed approach is compared to methods that use labeled data.

\subsubsection{Bilateration and Bounding-Box (BBB) Losses \cite{taner2025channel}}
This approach also enhances conventional CC to improve positioning performance without the use of labeled data and it is based on two assumptions. The first assumption is the knowledge of the AP positions. From the CSI matrix $\Hbm[a,n]$, the power of AP $a$ at $\tstamp{n}$ is computed as $P^{(a,n)}=20\log_{10}(\normFro{\Hbm[a,n]})$. This defines the set of estimated LoS APs for each UE position as
\begin{align} \label{eq:set_LoS_APs}
    \Aset^{(n)} = \{ a\in\Aset: P^{(a,n)} > P_{\mathrm{thr}}\},
\end{align}
where $P_{\mathrm{thr}}$ is a threshold set after observing the values of $P^{(a,n)}$ to ensure an effective distinction of \ac{LoS} and \ac{NLoS} samples.
From $\Aset^{(n)}$, they define the set of AP pairs 
\begin{align}
    \Psetn = \{(\apc, \apf)\in (\Aset^{(n)})^2:P^{(\apc,n)}>P^{(\apf,n)}+M_{\mathrm{p}}\},
\end{align}
where $P^{(\apc,n)}$ and $P^{(\apf,n)}$ are the power received by a close and far AP, respectively, and $M_{\mathrm{p}}$ ensures that only pairs of APs whose powers differ by at least $M_{\mathrm{p}}$ are included. The set $\Psetn$ is the foundation for the \textit{bilateration loss}, defined as
\begin{align} \label{eq:loss_taner_bi}
    \Lfrak[bi](\thetabm) = &\frac{1}{\sum_{n\in\Nset} |\Psetn|} \sum_{n\in\Nset} \sum_{(\apc, \apf)\in\Psetn} \nonumber \\
    &(\norm{\xbmhat[n]-\xbmunder[\apc]} - \norm{\xbmhat[n]-\xbmunder[\apf]}+M_{\mathrm{b}})^+,
\end{align}
where $\xbmunder[a]$ is the location of AP $a$ and $M_{\mathrm{b}} \geq 0$ is a hyper-parameter that ensures that $\xbmhat[n]$ is at least $M_{\mathrm{b}}$ closer to $\xbmunder[\apc]$ than to $\xbmunder[\apf]$. The second assumption is the knowledge of a bounding box corresponding to the LoS area of each AP. The bounding box for AP $a$ is defined as
\begin{align} \label{eq:def_bounding_box}
    \Bset^{(a)} = \{(x,y)\in\realset[2]: x\in [x_{\min}^{(a)}, x_{\max}^{(a)}], y\in[y_{\min}^{(a)}, y_{\max}^{(a)}]\},
\end{align}
where the values $x_{\min}^{(a)}, x_{\max}^{(a)}, y_{\min}^{(a)}, y_{\max}^{(a)}$ are set based on geometrical information of the environment.
This assumption enables the definition of a second loss function as
\begin{align} \label{eq:loss_taner_box}
    \Lfrak[box](\thetabm) = \frac{1}{\sum_{n\in\Nset}|\Aset^{(n)}|}\sum_{n\in\Nset} \sum_{a\in\Aset^{(n)}} l_a(\xbmhat[n]),
\end{align}
with 
\begin{align} \label{eq:loss_taner_box_la}
    l_a(\xbm) = \mathds{1}\{\xbm\notin \Bset^{(a)}\} \min_{\xbm'\in\Bset^{(a)}} \norm{\xbm-\xbm'}^2.
\end{align}
This loss enforces estimated positions $\xbmhat[n]$ in the LoS area of AP $a$ to belong to the bounding box of that AP.
Note that~\eqref{eq:loss_taner_box} still relies on the set $\Aset^{(n)}$ since the estimated positions can lie in arbitrary coordinates, and the received signals must be associated with the corresponding \ac{LoS} \acp{AP}.
The losses in \eqref{eq:loss_taner_bi} and \eqref{eq:loss_taner_box} are combined with the timestamp-based triplet-loss in \eqref{eq:triplet_loss} as 
\begin{align} \label{eq:bbb_loss}
    \Lfrak[BBB](\thetabm) = \lambdaCC \Lfrak[CC](\thetabm) + \lambdabi \Lfrak[bi](\thetabm) + \lambdabox\Lfrak[box](\thetabm).
\end{align}
Compared to \cite{taner2025channel}, we assume that a DT provides large-scale features and we do not require thresholding the power received by the APs to discern if the UE is in the LoS area.\footnote{ \label{foot:dt}One can also leverage a DT to compute the set $\Aset^{(n)}$ in~\eqref{eq:set_LoS_APs} based on synthetic data. However, the DT and the real scenario may be mismatched. Thus, we follow the thresholding of~\cite{taner2025channel} based on real data.}

\subsubsection{Affine Transformation with Ground-truth Positions}
This baseline considers a two-step approach for positioning. First, the loss in \eqref{eq:total_loss} with $\lambdaCC=1, \lambdadt=0$ is used to train the NN, i.e., we use the timestamp-based triplet loss and the predefined positions to obtain a first estimate $\bar{\xbm}^{(n)}$ of the positions. Then, we assume that ground-truth positions are available for the entire training dataset and we solve the following least-squares problem
\begin{align}
    \{\hat{\Abm}, \hat{\bbm}\} = \arg\min_{\substack{\Abm\in\realset[\Dx][\Dx]\\ \bbm\in\realset[\Dx]}} \sum_{n\in\Nset} \norm{(\Abm \bar{\xbm}^{(n)} + \bbm)-\xbm[n]}^2,
\end{align}
with $\xbm[n]$ the true UE position at time~$\tstamp{n}$. The estimated position is computed using the affine transform~$\xbmhat[n] = \hat{\Abm} \bar{\xbm}^{(n)} + \hat{\bbm}$. This method assumes a labeled data collection step prior to training the \ac{NN}. This baseline represents an extreme case of methods that use affine transformations, e.g., \cite{pihlajasalo2020absolute, euchner2024uncertainty}, where all the samples are used to compute the optimal affine transformation.

\subsubsection{CSI Fingerprinting~\cite{he2016wifi, wang2017csi}}
In CSI fingerprinting, labeled data of the true position of the \ac{UE} is also available and it consists of two phases. In the first phase, known as the offline training phase, the labeled dataset $\{\Hbm[n], \xbmn\}_{n=1}^N$ relating CSI and UE positions is used to train the positioning function $\gtheta{\cdot}$. To arrive at a fair comparison with our method, the position is estimated following the modified CC pipeline of our proposed method, i.e., $\xbmhat[n] = q(\pbm[n])$. The loss used in fingerprinting during training is 
\begin{align} \label{eq:sl_loss}
    \Lfrak[SL](\thetabm) = \frac{1}{N} \sum_{n\in\Nset} \norm{\xbmn - \xbmhat[n]}^2.  
\end{align} 
In the second phase, known as the online test phase, the learned positioning function is used to estimate the positions based on new observed CSI data.
CSI fingerprinting constitutes a lower bound in terms of positioning error since it uses ground-truth positions and unlike the affine transformation, the true positions are included in the loss function to train the NN, providing a more general mapping between the features $\fbm[n]$ and the estimated positions $\xbmhat[n]$.

\subsection{Scenario} \label{subsection:scenario}
\begin{figure}[tb]
    \centering
    \includestandalone[width=0.5\textwidth]{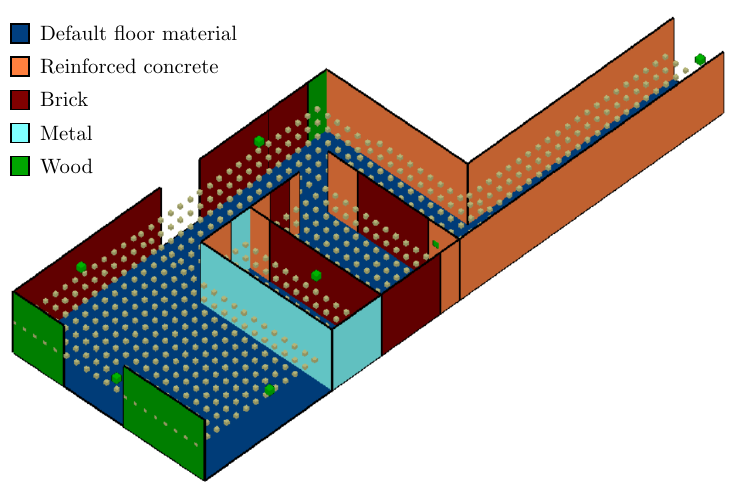}
    \caption{Simulated indoor scenario. The light green points represent the AP positions and the beige points represent the predefined points in the DT. The colors of the walls represent different materials, where the default floor material is provided by Wireless InSite.}
    \label{fig:sim_scenario}
\end{figure}

\begin{table}[tb]
\centering
\caption{Parameters of the simulated indoor scenario.}
\label{tab:sim_scenario_params}
\begin{tabular}{@{}ll@{}}
\toprule
Parameter & Value \\ \midrule
Number of APs & $A=8$ \\
Number of antennas per AP & $K=4$ \\
Number of antennas of the UE & $L=1$ \\
AP, UE and DT antenna type & Half-wavelength dipole \\
ULA spacing & Half-wavelength \\
Number of UE positions & $N=14606$ \\
Number of DT positions & $P = 713$ \\
Spacing between UE positions & 2.5 cm \\
Spacing between DT positions & $\Delta=0.5$ m \\
AP antenna height & 2 m \\
UE antenna height & 1.5 m \\
DT positions height & 1.5 m \\
Carrier frequency & 2.4 GHz \\
Bandwidth & 20 MHz \\
Number of subcarriers & $S=52$ \\
Number of truncated taps & $C=13$ \\ 
Maximum SNR per AP & 25 dB \\ \midrule
Learnable function $\gtheta{\cdot}$ & Multi-layer perceptron \\
Number of hidden layers & 4 \\
Neurons per hidden layer & 256 \\ 
Hidden layer activation function & Rectified linear unit (ReLU) \\
Optimizer & Adam \\
\multirow{2}{*}{Learning rate} & $5\cdot10^{-3}$, \textit{APP}, \textit{TDP} \\
& $1\cdot10^{-3}$, otherwise \\
Training iterations & $5000$ \\
Dropout (hidden layers) & 0.15 \\
Coherence time in \eqref{eq:T_set} & $\Tc=\Tf=2$ s \\
Distance threshold in \eqref{eq:triplet_loss} & $\Mt=0.9$ m \\
\multirow{3}{*}{CC weight in \eqref{eq:total_loss}} & $\lambdaCC=10$, \textit{Power} and \textit{APP} \\ 
& $\lambdaCC=5$, \textit{DPP} and \textit{Covariance} \\ 
& $\lambdaCC=1$, \textit{TDP} \\
\bottomrule
\end{tabular}
\end{table}

We simulate the uplink D-MIMO indoor scenario of \cite{taner2025channel}, illustrated in Fig.~\ref{fig:sim_scenario}. In our simulations, we particularize the proposed framework in Sec.~\ref{subsec:proposed_method} for a two-dimensional position estimation problem, i.e., $\Dx=2$ and a single UE in the environment. The UE moves in this scenario (see Fig.~\ref{fig:pos_sim_results_true} for a view of the trajectory of the UE), covering all rooms, which defines the training and testing sets. We consider a grid of points as the predefined positions in the DT, represented in Fig.~\ref{fig:sim_scenario}.\footnote{\label{foot:large_areas}In this work, we assume that a single DT covers the entire scenario. Very large environments require division into smaller zones and subsequent zone stitching or identification, which is left as a future work.} The simulations use Remcom's Wireless InSite ray-tracing software to obtain the CSI matrices $\{\Hbm[n]\}_{n=1}^N$ and the simulated CSI $\{\Hbmtilde[p]\}_{p=1}^P$. 
Innitially, there is a perfect match between the DT and the real scenario (see Sec.~\ref{subsec:robustness} for the mismatched case).
The \textit{Power} feature is directly available from the software, whereas the \textit{APP, DPP}, and \textit{Covariance} features are derived by first simulating the CSI matrices $\{\Hbmtilde[p]\}_{p=1}^P$ and applying the functions described in Sec.~\ref{subsec:features} to compute the DT features $\vbmbar[p]$. 
Since no real measurements are available, we contaminate the CSI matrices $\Hbm[n]$ with \ac{AWGN}. We define the SNR for AP $a$ and timestamp $\tstamp{n}$ as $\SNR^{(a, n)}=\norm{\hbm[a,n]}^2/\sigma^2$, with $\hbm[a,n]=\vecrm{[\Hbm[a,n]\dft{S}^{\hermit}]_{:,1:C}}$ and $\sigma^2$ the variance of the AWGN. We add AWGN to $\Hbm[n]$ such that $\max_{a,n}\{\SNR^{(a,n)}\}=25$ dB across training and testing sets. 
To compare with the results of \cite{taner2025channel}, we do not add AWGN to samples where $\Hbm[a,n]=0$, but these samples are still used for the computation of the CSI features $\fbm[n]$ and the large-scale features $\vbm[n]$. 
Note that we do not add \ac{AWGN} to the DT CSI matrices $\Hbmtilde[p]$.
The dataset $\{\Hbm[n]\}_{n=1}^N$ is randomly split into training and test sets, with 80\% of samples used for training.\footnote{\label{foot:num_samples}In hardware-limited systems, the number of training samples can be reduced, as shown in~\cite{taner2023streaming}. However, this study focuses on the best attainable positioning performance for the given training dataset.} As learnable function for positioning $\gtheta{\fbm[n]}$, we consider a multi-layer perceptron with input and output dimensions determined by $\Dh$ in \eqref{eq:input_nn} and $P$, respectively. The learnable parameters $\thetabm$ correspond to the weights and biases of the MLP layers. The simulation parameters are described in Table~\ref{tab:sim_scenario_params}. The dataset of collected CSI from the real scenario is based on standard 5G requirements. For example, a recent 5G testbed captured CSI with a target SNR of 28 dB every 10--20 ms~\cite{wiesmayr2025csi}. A spatial resolution of 2.5 cm would correspond to an average UE velocity of 1.25--2.5 m/s, which lies within the typical walking speed range of healthy individuals~\cite{BOHANNON2011182}. Moreover, the considered unlabeled dataset would be collected in less than five minutes and the proposed approach can be executed for prolonged durations.

\subsection{Positioning Results} \label{subsec:sim_results}
\begin{table*}[tb]
\centering
\caption{Positioning results for the simulated indoor scenario.}
\label{tab:sim_results}
\begin{tabular}{@{}lllcccccc@{}}
\toprule
 & Method & Figure & TW $\uparrow$ & CT $\uparrow$ & KS $\downarrow$ & RD $\downarrow$ & MDE $\downarrow$ & 95th PDE $\downarrow$ \\ \midrule
\multirow{3}{*}{Baselines} & BBB \cite{taner2025channel} & \ref{fig:pos_sim_results_sueda} & $0.979$ & $0.976$ & $0.144$ & $0.710$ & $1.15\pm0.04$ & $2.84\pm0.44$ \\
& Affine transformation & \ref{fig:pos_sim_results_affine} & $0.954$ & $0.957$ & $0.375$ & $0.879$ & $3.20\pm0.64$ & $6.53\pm1.43$ \\
& CSI fingerprinting & \ref{fig:pos_sim_results_finger}  & $0.993$  & $0.993$  & $0.060$  & $0.479$  &  $0.41\pm0.02$ & $1.01\pm0.08$ \\ \midrule
\multirow{5}{*}{Proposed} & \textit{Power} & \ref{fig:pos_sim_results_pow} & $0.990$ & $\bm{0.989}$ & $\bm{0.098}$  & $\bm{0.603}$  & $\bm{0.81\pm0.05}$ & $\bm{1.78\pm0.13}$  \\
 & \textit{Angle-power profile (APP)} & \ref{fig:pos_sim_results_angle} & $\bm{0.991}$  & $0.988$  & $0.111$ & $0.631$ & $1.06\pm0.07$ & $2.67\pm0.34$ \\
 & \textit{Delay-power profile (DPP)} & \ref{fig:pos_sim_results_delay} & $0.990$  & $0.987$  & $0.105$ & $0.620$ & $0.91\pm0.07$ & $2.27\pm0.36$ \\
 & \textit{Covariance} & \ref{fig:pos_sim_results_cov} &  $\bm{0.991}$ & $0.985$  & $0.126$  & $0.658$  & $1.15\pm0.14$ & $3.11\pm0.45$ \\ 
 &  \textit{Truncated delay profile (TDP)} & \ref{fig:pos_sim_results_tdp} & $0.989$ & $0.982$  & $0.143$  & $0.665$  & $1.24\pm0.46$ & $3.63\pm2.47$ \\
\bottomrule
\end{tabular}
\end{table*}

\begin{figure*}[tb]
    \centering

    % Top mini-figure (legend)
    \includegraphics[width=0.15\textwidth]{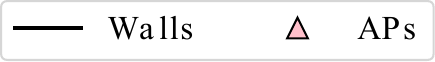}\\[0.5ex]  % adjust the width and spacing here

    %Subfigures
    \begin{subfigure}{0.19\textwidth}
        \centering
        \includegraphics[width=\textwidth]{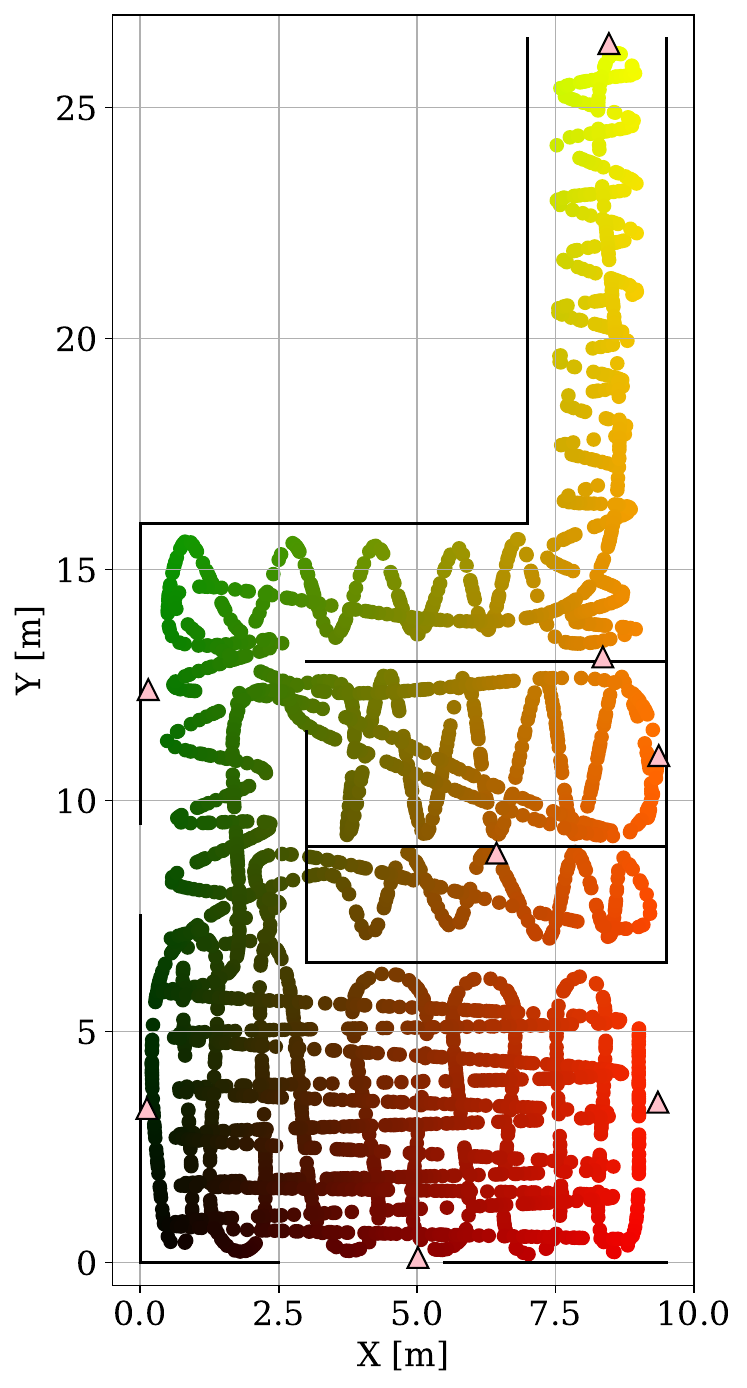}
        \caption{True trajectory.}
        \label{fig:pos_sim_results_true}
    \end{subfigure}
    \begin{subfigure}{0.19\textwidth}
        \centering
        \includegraphics[width=\textwidth]{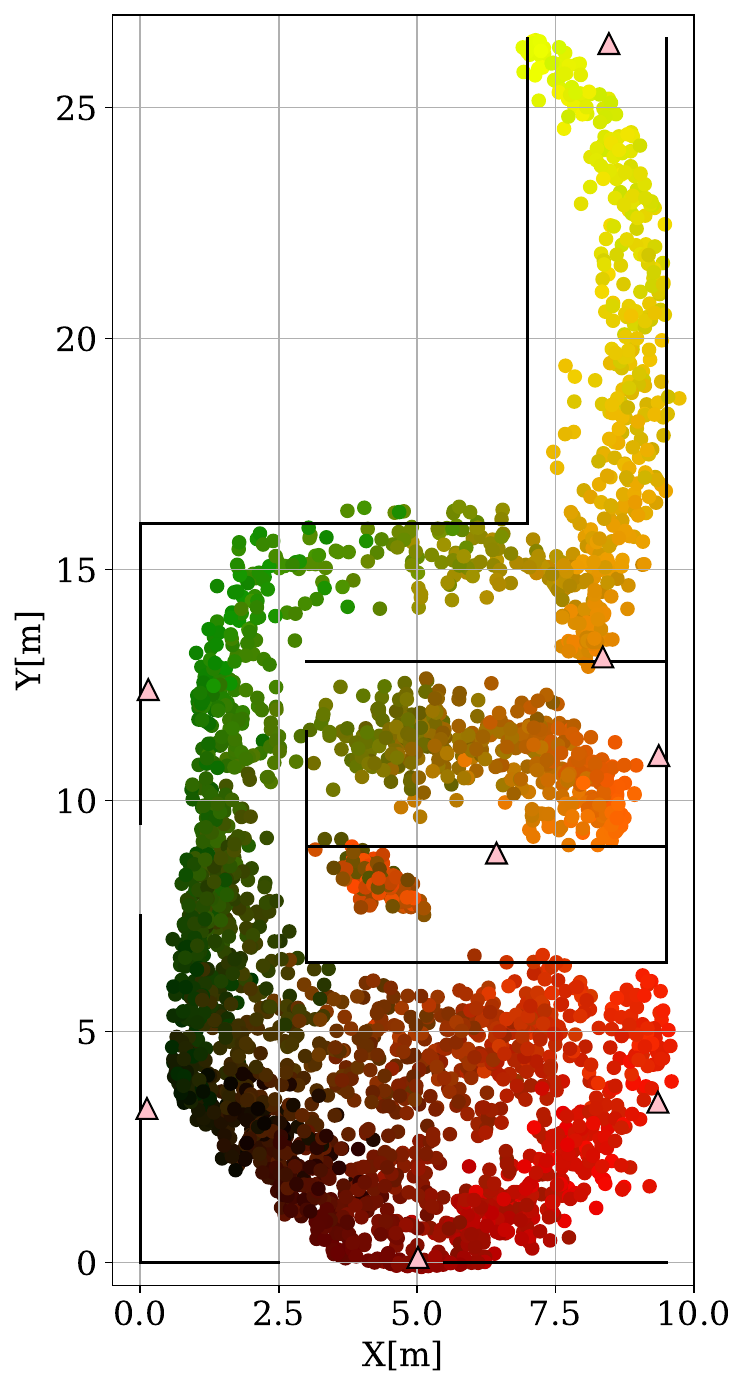}
        \caption{BBB \cite{taner2025channel}.}
        \label{fig:pos_sim_results_sueda}
    \end{subfigure}
    \begin{subfigure}{0.19\textwidth}
        \centering
        \includegraphics[width=\textwidth]{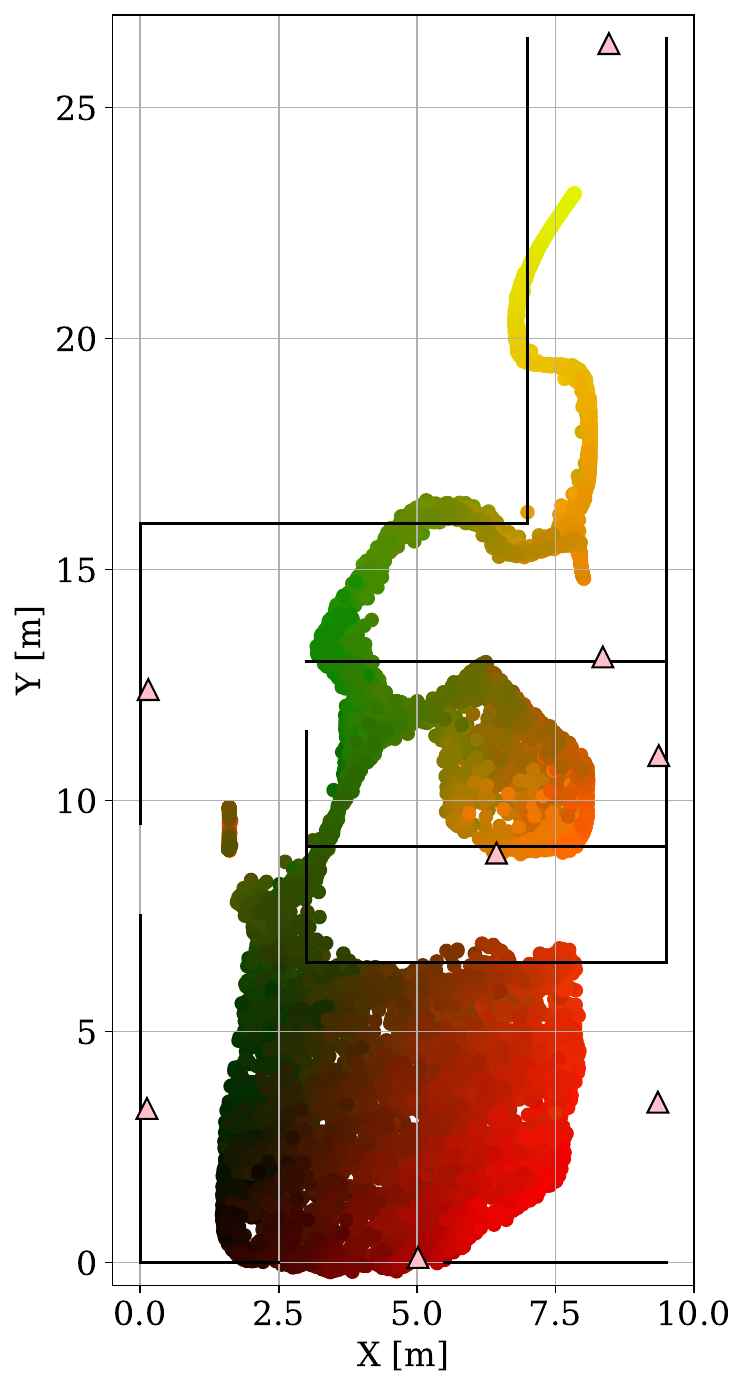}
        \caption{Affine transformation.}
        \label{fig:pos_sim_results_affine}
    \end{subfigure}
    \begin{subfigure}{0.19\textwidth}
        \centering
        \includegraphics[width=\textwidth]{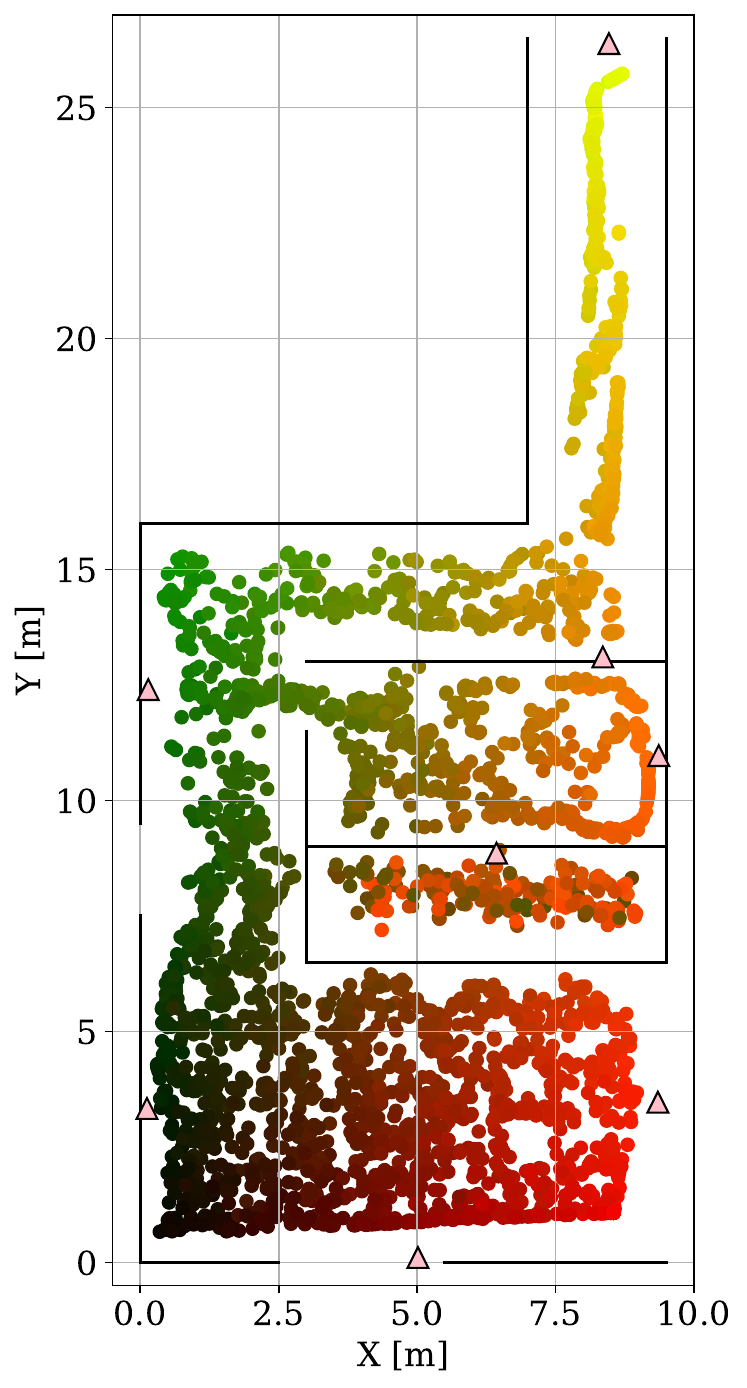}
        \caption{CSI fingerprinting.}
        \label{fig:pos_sim_results_finger}
    \end{subfigure}

    \par\medskip
    
    \begin{subfigure}{0.19\textwidth}
        \centering
        \includegraphics[width=\textwidth]{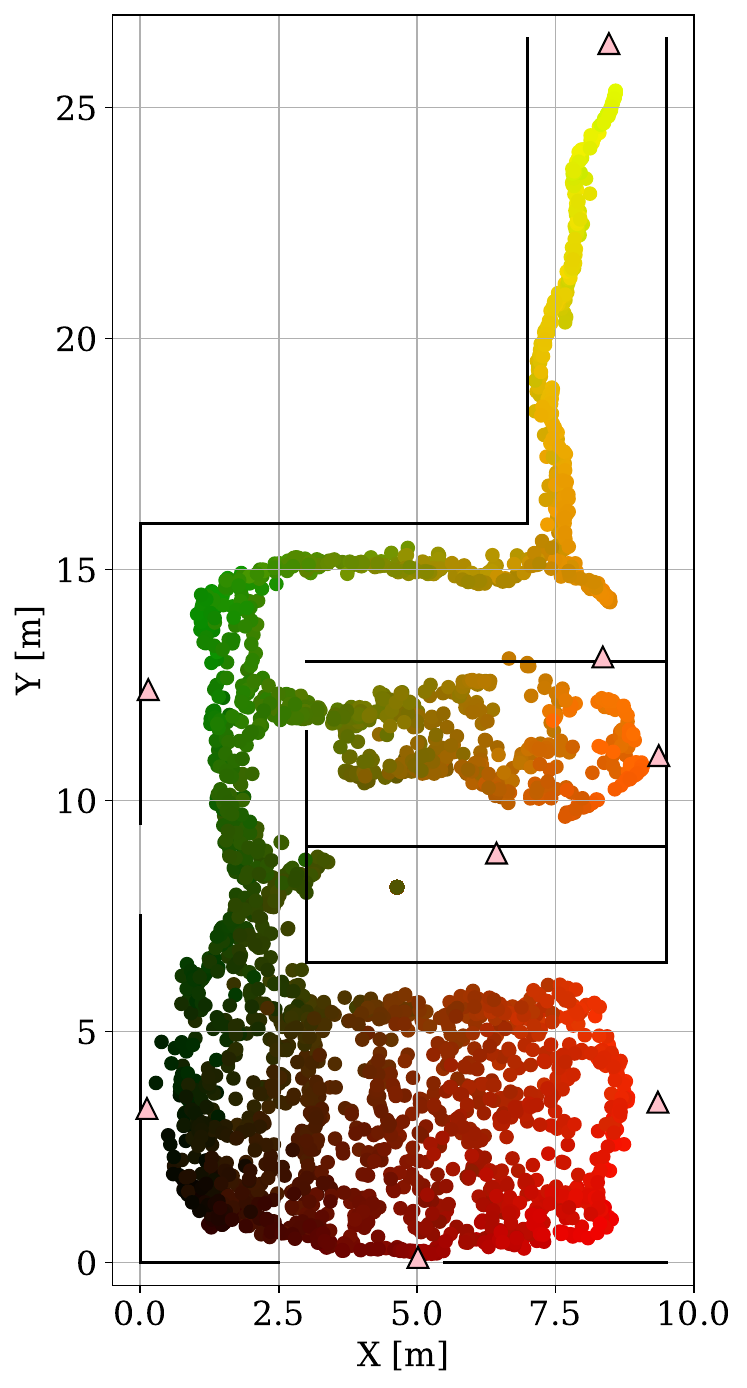}
        \caption{\textit{Power}.}
        \label{fig:pos_sim_results_pow}
    \end{subfigure}
    \begin{subfigure}{0.19\textwidth}
        \centering
        \includegraphics[width=\textwidth]{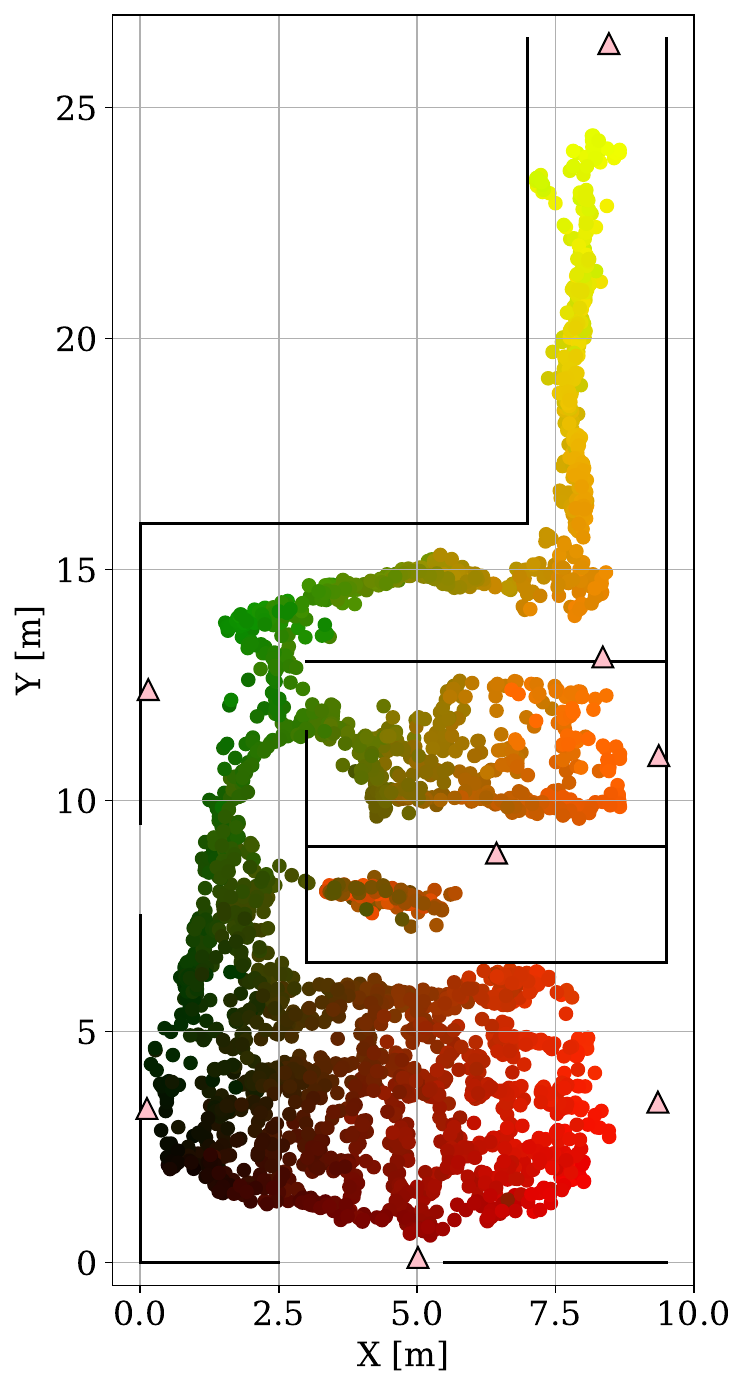}
        \caption{\textit{Angle-power profile.}}
        \label{fig:pos_sim_results_angle}
    \end{subfigure}
    \begin{subfigure}{0.19\textwidth}
        \centering
        \includegraphics[width=\textwidth]{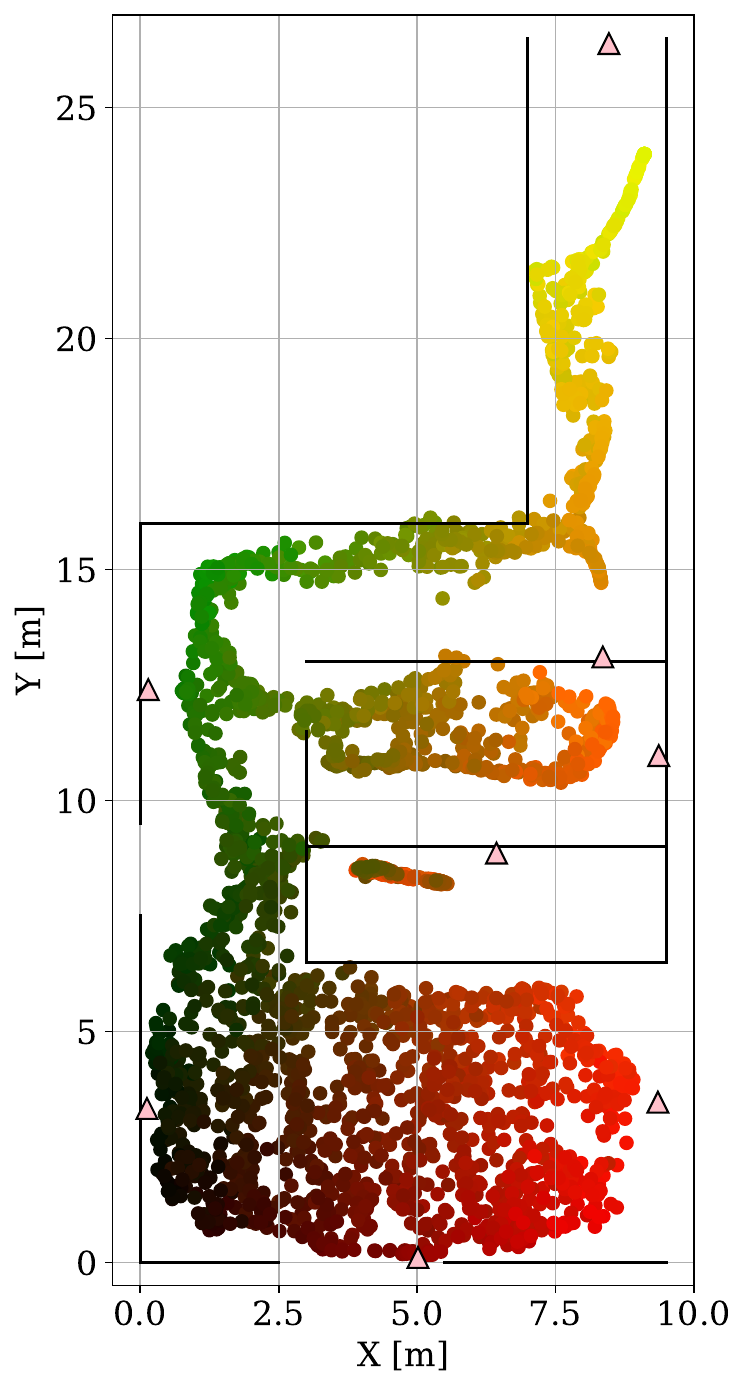}
        \caption{\textit{Delay-power profile.}}
        \label{fig:pos_sim_results_delay}
    \end{subfigure}
    \begin{subfigure}{0.19\textwidth}
        \centering
        \includegraphics[width=\textwidth]{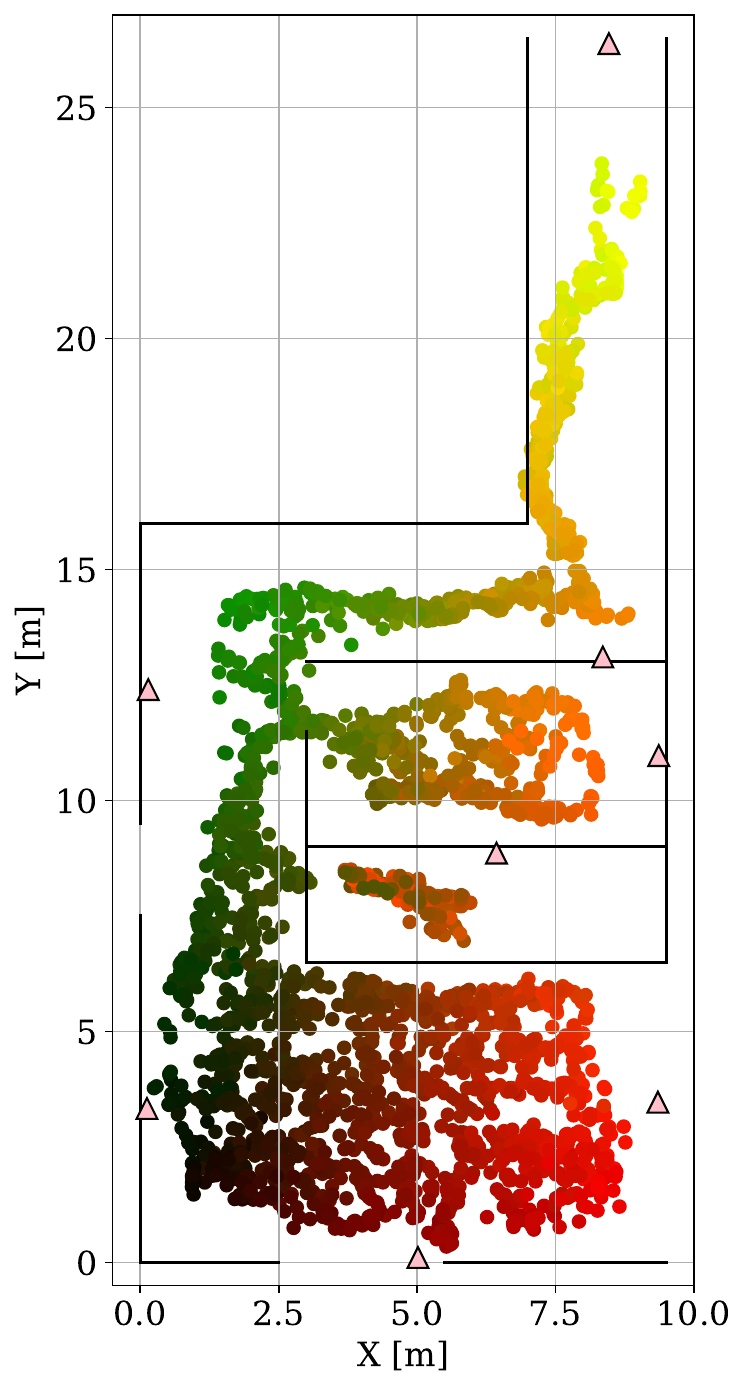}
        \caption{\textit{Covariance.}}
        \label{fig:pos_sim_results_cov}
    \end{subfigure}
    \begin{subfigure}{0.19\textwidth}
        \centering
        \includegraphics[width=\textwidth]{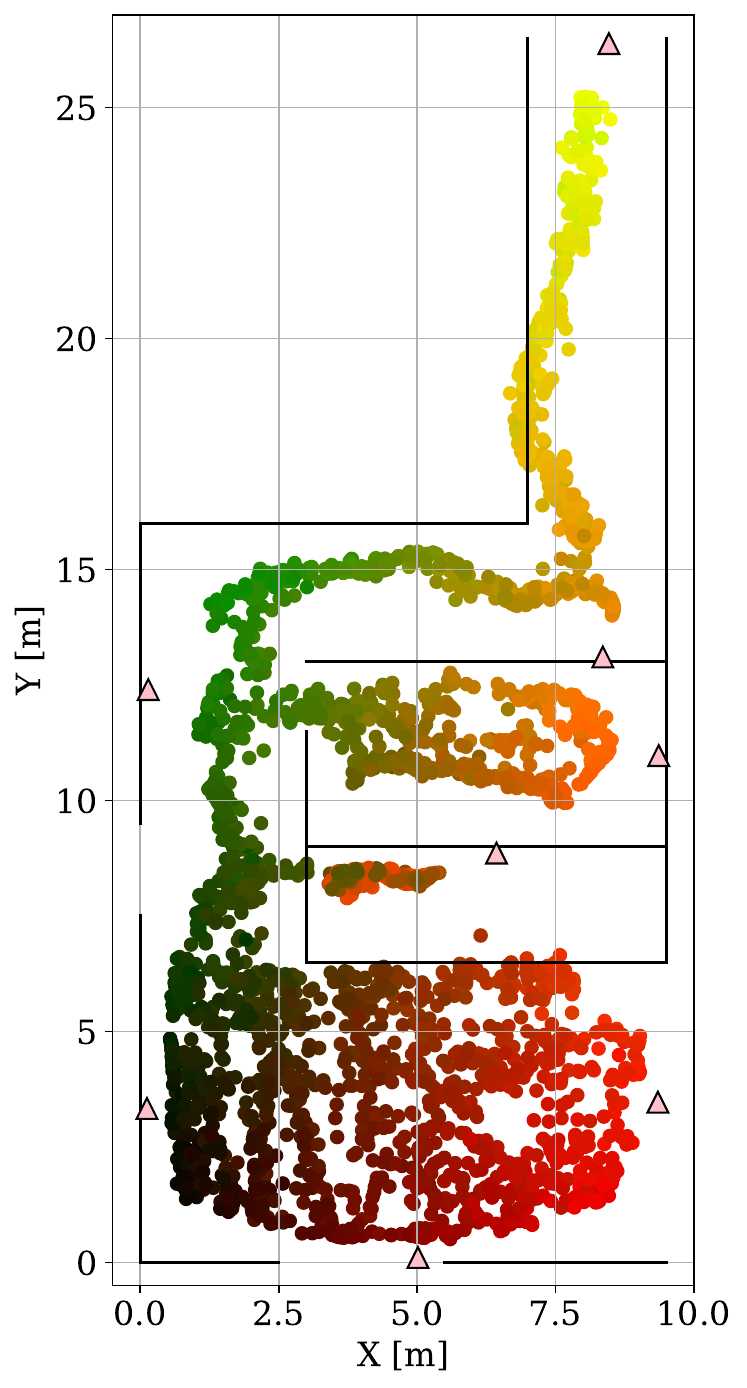}
        \caption{\textit{Truncated Delay Profile.}}
        \label{fig:pos_sim_results_tdp}
    \end{subfigure}
    \caption{True and estimated trajectories for the proposed approaches and the baselines in the simulated indoor scenario. For each estimated trajectory, we represent the best of the 10 realizations of the NN.}
    \label{fig:pos_sim_results}
\end{figure*}

Table~\ref{tab:sim_results} presents the inference results for the indoor scenario in Fig.~\ref{fig:sim_scenario}. The metrics are described in Appendix~\ref{app:eval_metric} and the arrows next to the evaluation metrics in Table~\ref{tab:sim_results} indicate whether a higher or lower value is better. For practical applications, the \ac{MDE} is the foremost metric to evaluate, but other metrics are included for completeness. The results are averaged over $10$ random NN initializations and expressed as $\mu\pm\sigma$, where $\mu$ and $\sigma$ are the mean and standard deviation, respectively. The standard deviations for \ac{TW}, \ac{CT}, \ac{KS}, and \ac{RD} are omitted due to their negligible values (similar to \cite{taner2025channel}).
The results in Table~\ref{tab:sim_results} show that the proposed \textit{Power}, \textit{APP}, and \textit{DPP} features outperform the state-of-the-art method in \cite{taner2025channel} and the affine transformation. This indicates that the use of a DT improves positioning performance compared to (i) the knowledge of the AP positions and their LoS area and (ii) methods that rely on the timestamp-based triplet loss and affine transformations. As expected, fingerprinting achieves superior performance, but it requires labeled UE position data. 

Among the proposed features, \textit{Power} performs best for most evaluation metrics, which suggests that relying on smooth large-scale features over space provides better results. The performance degrades when small-scale features are introduced in the \textit{APP}, \textit{DPP}, \textit{TDP}, and \textit{Covariance} features, confirming the relevance of large-scale features for CC \cite{studer2018channel}.

Fig.~\ref{fig:pos_sim_results} shows a top-down view of the true and estimated UE trajectories according to the proposed features and the baselines listed in Table~\ref{tab:sim_results}. As shown in Proposition~\ref{prop:power}, the \textit{Power} feature performs poorly in the area enclosed by four walls, where only a single AP is available. The \textit{APP} and \textit{DPP} features provide an alternative solution that improves the performance in such areas at the cost of a slightly worse overall performance.
The results in Table~\ref{tab:sim_results} and Fig.~\ref{fig:pos_sim_results} indicate that the proposed approach is an effective CSI-based positioning method without requiring labeled data. Given that the \textit{TDP} and \textit{Covariance} features yield the poorest performances, they are not further considered in the remainder of the paper.

\subsubsection{Effect of the Grid Spacing in the DT}
In this test, we study the performance of the proposed approach as a function of the spacing $\Delta$ in the predefined DT positions. The spacing in Table~\ref{tab:sim_results} was $\Delta=0.5$ m, corresponding to the grid shown in Fig.~\ref{fig:sim_scenario}. In Table~\ref{tab:results_grid}, we show the MDE and 95th \ac{PDE} results for different spacings in the predefined DT positions. 
The NN architecture described in Table~\ref{tab:sim_scenario_params} remains unchanged across $\Delta$ values except in the last layer, which depends on the number of predefined points $P$.

The results show that a finer spacing of $\Delta=0.25$ m leads to a significant average performance drop and a larger standard deviation across NN initializations. However,  the best-performing random initializations for $\Delta=0.25$\,m yield MDE$~=~0.96$\,m and 95th \ac{PDE}$~=~2.52$\,m for \textit{Power}, MDE$~=~1.58$\,m and 95th \ac{PDE}$~=~4.66$\,m for \textit{APP}, and MDE$~=~0.79$\,m and 95th \ac{PDE}$~=~1.74$\,m for \textit{DPP} which are comparable to the average performance obtained for $\Delta>0.25$\,m. Given that the NN architecture and the learning hyper-parameters are optimized for $\Delta=0.5$\,m, changing the grid spacing $\Delta$ reduces the convergence stability of the optimization problem in \eqref{eq:proposed_problem} across different random initializations. 

The \textit{Power} feature obtains similar results for $\Delta>0.5$ m, while the \textit{APP} and \textit{DPP} features degrade in performance with a larger spacing. This difference arises because the \textit{Power} is a large-scale feature that fluctuates more slowly over space compared to the \textit{APP}, allowing the MLP to effectively interpolate feature vectors for off-grid positions. In contrast, the faster spatial variations of the \textit{APP} and \textit{DPP} hinder interpolation when the grid spacing is large. From the results in Table~\ref{tab:results_grid}, we conclude that the \textit{Power} feature is robust under a sparser grid of predefined points in the DT, enabling reduced DT complexity without substantial performance loss.

\begin{table}[tb]
\centering
\caption{Testing results for different position spacings $\Delta$ in the DT. The MDE (top) and 95th \ac{PDE} (bottom) are shown in the form of $\mu\pm\sigma$, with $\mu$ and $\sigma$ the mean and standard deviation over 10 random NN realizations.}
\label{tab:results_grid}
\begin{tabular}{@{}lcccc@{}}
\toprule
Method & $\Delta=0.25$ m & $\Delta=0.5$ m & $\Delta=1$ m & $\Delta=1.5$ m  \\ \midrule
\multirow{2}{*}{\textit{Power}} & $4.21\pm3.74$ & $0.81\pm0.05$ & $0.91\pm0.02$ & $0.83\pm0.03$ \\
& $9.05\pm6.59$ & $1.78\pm0.13$ &  $1.90\pm0.08$ & $1.80\pm0.11$\\ \midrule
\multirow{2}{*}{\textit{APP}} & $5.32\pm2.99$ & $1.06\pm0.07$  & $1.26\pm0.05$ & $1.66\pm0.25$ \\
 & $11.02\pm4.67$ & $2.67\pm0.34$  & $2.64\pm0.28$ & $3.38\pm0.84$ \\ \midrule
\multirow{2}{*}{\textit{DPP}} & $3.74\pm3.33$ & $0.91\pm0.07$  & $1.18\pm0.18$ & $1.72\pm1.72$ \\
 & $7.73\pm5.88$ & $2.27\pm0.36$  & $2.67\pm0.93$ & $4.00\pm3.25$ \\
 \bottomrule
\end{tabular}
\end{table}

\subsubsection{Effect of Labeled Data on Fingerprinting} \label{subsubsec:few_labeled_data}
\begin{figure}[tb]
    \centering
    \includestandalone[width=0.47\textwidth]{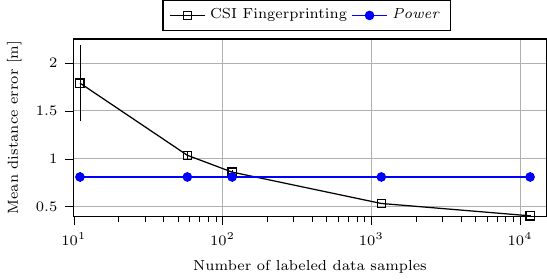}
    \caption{Positioning performance as a function of the number of labeled data samples during training. The curves indicate the mean and the error bars the standard deviation across random NN initializations.}
    \label{fig:results_fewer_labels}
\end{figure}

The results in Table~\ref{tab:sim_results} show a substantial performance difference between the proposed approach and fingerprinting. However, fingerprinting leverages the entire database as labeled samples, which is typically costly to obtain. 
To reduce the costs of fingerprinting, we combine the \ac{CC} loss of~\eqref{eq:triplet_loss} and the supervised loss of~\eqref{eq:sl_loss} as 
\begin{align} \label{eq:sl_cc_loss}
    \Lfrak(\thetabm) = \lambdaCC\Lfrak[CC] + \lambda_{\mathrm{SL}}\Lfrak[SL],
\end{align}
where $\lambda_{\mathrm{SL}}>0$ is a hyper-parameter that balances the contribution of the \ac{CC} loss and the supervised loss. Moreover, we reduce the number of labeled training samples used for fingerprinting and compare the resulting performance with the proposed approach. 
In Fig.~\ref{fig:results_fewer_labels}, we report the \ac{MDE} of fingerprinting following~\eqref{eq:sl_cc_loss}, together with the proposed approach based on the \textit{Power} feature. The training samples for fingerprinting are uniformly distributed in space, and we set $\lambdaCC=1$ and $\lambda_{\mathrm{SL}}=10$. Our proposed approach is label-agnostic, resulting in the flat behavior observed in Fig.~\ref{fig:results_fewer_labels}. The performance of fingerprinting degrades with fewer labeled samples since their spatial coverage becomes sparser and extrapolation to unseen locations becomes more challenging.
Additionally, under 100 labeled samples the proposed approach using the \textit{Power} feature outperforms fingerprinting. This result underscores the dependence of fingerprinting on dense labeled datasets and its higher data acquisition cost compared to the proposed approach.

\subsection{Robustness Tests} \label{subsec:robustness}
In this section, we evaluate the robustness of the proposed approach against mismatches at training and inference time. The former affects the assumed model of the DT while the latter studies the effect of unexpected elements in the real environment once the proposed approach has been trained.

\begin{figure}[tb]
    \centering
    \includestandalone[width=0.5\textwidth]{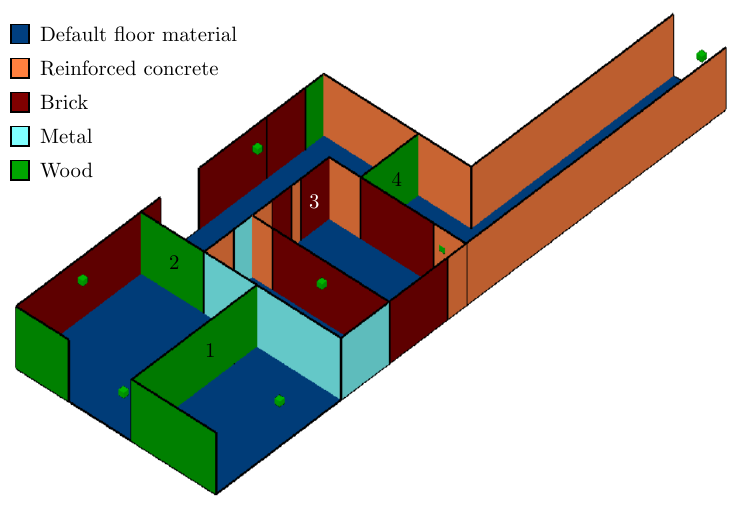}
    \caption{Mismatched digital twin in the case of four additional walls. The additional walls are indicated with numbers. The case with \textit{one additional wooden wall} only considers wall number 4.}
    \label{fig:mismatched_dt}
\end{figure}
\subsubsection{Mismatched Digital Twin}\label{subsec:mismatched_dt}
\begin{table*}[b]
\centering
\caption{Mismatched DT test results. The MDE (top) and 95th \ac{PDE} (bottom) are shown in the form of $\mu\pm\sigma$, with $\mu$ and $\sigma$ the mean and standard deviation over 10 random NN realizations.}
\label{tab:robustness}
\begin{tabular}{@{}lccccc@{}}
\toprule
Method & No mismatch & \begin{tabular}[c]{@{}c@{}}Shifted \\ AP positions\end{tabular} & \begin{tabular}[c]{@{}c@{}}Material \\ mismatches\end{tabular} & \begin{tabular}[c]{@{}c@{}}One additional \\ wooden wall \end{tabular} & \begin{tabular}[c]{@{}c@{}}Four additional \\ walls\end{tabular} \\ \midrule 
\multirow{2}{*}{\textit{Power}} & $0.81\pm0.05$   & $0.83\pm0.06$ & $0.92\pm0.06$ & $0.91\pm0.04$ & $1.08\pm0.21$ \\
& $1.78\pm0.13$ & $2.02\pm0.40$ & $2.33 \pm 0.29$ & $2.20\pm 0.20$ & $2.49\pm0.75$ \\ \midrule
DT-based CSI & $0.41 \pm 0.02$  & $0.52\pm0.02$ & $0.65\pm 0.02$ & $0.88\pm0.06$ & $1.92\pm0.18$ \\
 fingerprinting & $1.01 \pm 0.08$ & $1.26\pm0.08$ & $1.65\pm 0.06$ & $2.44\pm0.31$ & $4.56\pm0.46$\\ \midrule
 DT-based \textit{Power} & $0.90\pm0.06$ & $0.99\pm0.04$ & $1.04\pm0.02$ & $1.15 \pm 0.06$ & $1.54\pm0.05$ \\
 fingerprinting & $2.20\pm0.15$ & $2.38\pm0.06$ & $2.52\pm0.04$ & $2.77\pm0.17$ & $3.45\pm0.10$ \\ \bottomrule
\end{tabular}
\end{table*}
So far, we have considered a perfect match between the DT and the considered scenario. In practice, information about the environment may be incomplete or inaccurate. We therefore evaluate the performance of the proposed approach under DT modeling mismatches during training. Particularly, we consider the following increasingly severe mismatches: (i) the APs in the DT are shifted by half a wavelength ($6.25$\,cm) in the positive direction of the X-axis, (ii) the walls and floor materials in the DT are changed to wood, (iii) an additional wooden wall is introduced in the DT, and (iv) four additional walls are introduced in the DT. The two last cases are represented in Fig.~\ref{fig:mismatched_dt}. 
Furthermore, given access to a DT, we consider two practical DT-based fingerprinting baselines: (i) training a fingerprinting model using synthetic CSI generated by the DT~\cite{morais2024localization}, and (ii) performing fingerprinting using synthetic \textit{Power} features from the DT. The latter corresponds to leveraging a computationally efficient DT~\cite{ait2024fast} and the NN architecture is accordingly adjusted, reducing the input layer to a single neuron. For both baselines, we use the training data described in Sec.~\ref{subsection:scenario}, i.e., the full training dataset. Table~\ref{tab:robustness} shows the inference results.

When the APs are shifted in the DT, the MDE of the proposed approach and DT-based \textit{Power} fingerprinting are similar to the case of no mismatches in the DT, while DT-based CSI fingerprinting shows a slight performance degradation. This result indicates that a displacement of half a wavelength substantially alters the phase of the estimated CSI, thereby introducing a mismatch between the training and testing fingerprinting data. In contrast, the large-scale \textit{Power} feature remains largely unchanged under the same displacement.
As we introduce more severe impairments, all methods generally degrade in performance, reflecting the growing discrepancy between the DT and the real scenario. However, the degradation trends differ across methods, as discussed next.

Comparing DT-based CSI fingerprinting with the proposed approach, the former achieves better performance under no or slight mismatches in the DT as it exploits the full CSI information from the DT, while the proposed approach relies on lower-dimensional large-scale features such as the received power. 
Under severe impairments (e.g., four additional wooden walls), the proposed approach outperforms DT-based CSI fingerprinting. This indicates two relevant aspects. First, large-scale features are more robust to modeling mismatches as they vary more smoothly over space. This observation is consistent with the fact that DT-based \textit{Power} fingerprinting outperforms DT-based CSI fingerprinting under severe mismatches. Second, incorporating CC in the proposed approach aligns the DT and the observed data, improving robustness. Conversely, DT-based CSI fingerprinting only harnesses synthetic fine-grained  data, which limits the generalization of the learned positioning function to the real environment. This is further supported by comparing DT-based \textit{Power} fingerprinting and the proposed approach, which use the same information from the DT, but the proposed framework additionally integrates observed data.

In summary, the results from Table~\ref{tab:robustness} indicate three regimes: (i) if a high-fidelity DT that computes the CSI is available, DT-based CSI fingerprinting yields the best results, at the cost of higher computational complexity; (ii) in the case of a low-fidelity (potentially mismatched) DT that generates CSI, DT-based CSI fingerprinting achieves better performance under slight mismatches, while the proposed approach remains 
robust under severe mismatches; and (iii) with a computationally efficient DT that only provides the received power, our approach achieves the best results.

\subsubsection{Distribution Shift}\label{subsubsec:distribution_shift}
\begin{figure}[tb]
    \centering
    \includestandalone[width=0.5\textwidth]{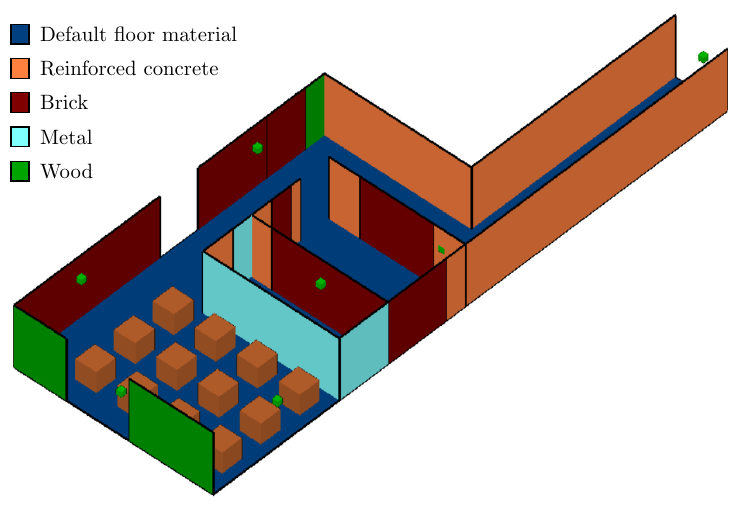}
    \caption{Mismatched scenario with additional wooden boxes in the large room. The boxes have a size of $1$ m in all axes.}
    \label{fig:mismatched_boxes}
\end{figure}
Finally, we evaluate the positioning performance under mismatches between training and testing data. 
Two types of distribution shifts are considered: (i) changes in the UE position and (ii) changes in the environmental model. Note that in both cases, the DT is matched to the real scenario during training.
For the first case, we consider that during
training, the estimated CSI corresponds to a UE at 1.5\,m height. At inference time, the estimated CSI corresponds to a UE at 0.8\,m height.
For the second case, we consider additional wooden boxes in the large room, as represented in Fig.~\ref{fig:mismatched_boxes}, or the same additional wooden wall as in Fig.~\ref{fig:mismatched_dt}. This simulation reflects practical scenarios where environmental changes occur after deployment (e.g., a door being closed).
The results are summarized in Table~\ref{tab:generalization}, where we also included the results from Table~\ref{tab:sim_results}. 

\begin{table}[tb]
\centering
\caption{Generalization test results. The MDE (top) and 95th \ac{PDE} (bottom) are shown in the form of $\mu\pm\sigma$, with $\mu$ and $\sigma$ the mean and standard deviation over 10 random NN realizations.}
\label{tab:generalization}
\begin{tabular}{@{}p{1.3cm}cc>{\color{black}}c>{\color{black}}c@{}}
\toprule
Method & No mismatch & \begin{tabular}[c]{@{}c@{}}Decreased \\ UE height\end{tabular} & \begin{tabular}[c]{@{}c@{}}Wooden \\ boxes\end{tabular} & \begin{tabular}[c]{@{}c@{}}Wooden \\ wall\end{tabular} \\ \midrule
\multirow{2}{*}{\textit{Power}} 
& $0.81\pm0.05$ & $0.97\pm0.05$ & $0.82\pm0.04$ & $1.10\pm0.08$ \\
& $1.78\pm0.13$ & $2.49\pm0.17$ & $1.80\pm0.14$ & $3.68\pm0.21$ \\ \midrule

\multirow{2}{*}{\textit{APP}} 
& $1.06\pm0.07$ & $1.26\pm0.06$ & $1.06\pm0.06$ & $1.43\pm0.05$  \\
& $2.67\pm0.34$ & $3.18\pm0.22$ & $2.69\pm0.34$  & $4.15\pm0.12$ \\ \midrule

\multirow{2}{*}{\textit{DPP}} 
& $0.91\pm0.07$ & $1.06\pm0.08$ & $0.91\pm0.07$ & $1.18\pm0.05$ \\
& $2.27\pm0.36$ & $2.70\pm0.35$ & $2.25\pm0.39$ & $3.82\pm 0.17$ \\ \midrule

CSI
& $0.41\pm0.02$ & $0.63\pm0.01$ & $0.43\pm0.01$ & $0.93\pm0.03$ \\ 
fingerprinting & $1.01\pm0.08$ & $1.79\pm0.04$ & $1.02\pm0.04$ & $3.79\pm0.23$ \\\bottomrule
\end{tabular}
\end{table}

\begin{figure}[tb]
    \centering
    % Top mini-figure (legend)
    \includegraphics[width=0.15\textwidth]{figures/results_sim/legend.pdf}\\[0.5ex]  % adjust the width and spacing here
    %Subfigures
    \begin{subfigure}{0.23\textwidth}
        \centering
        \includegraphics[width=\textwidth]{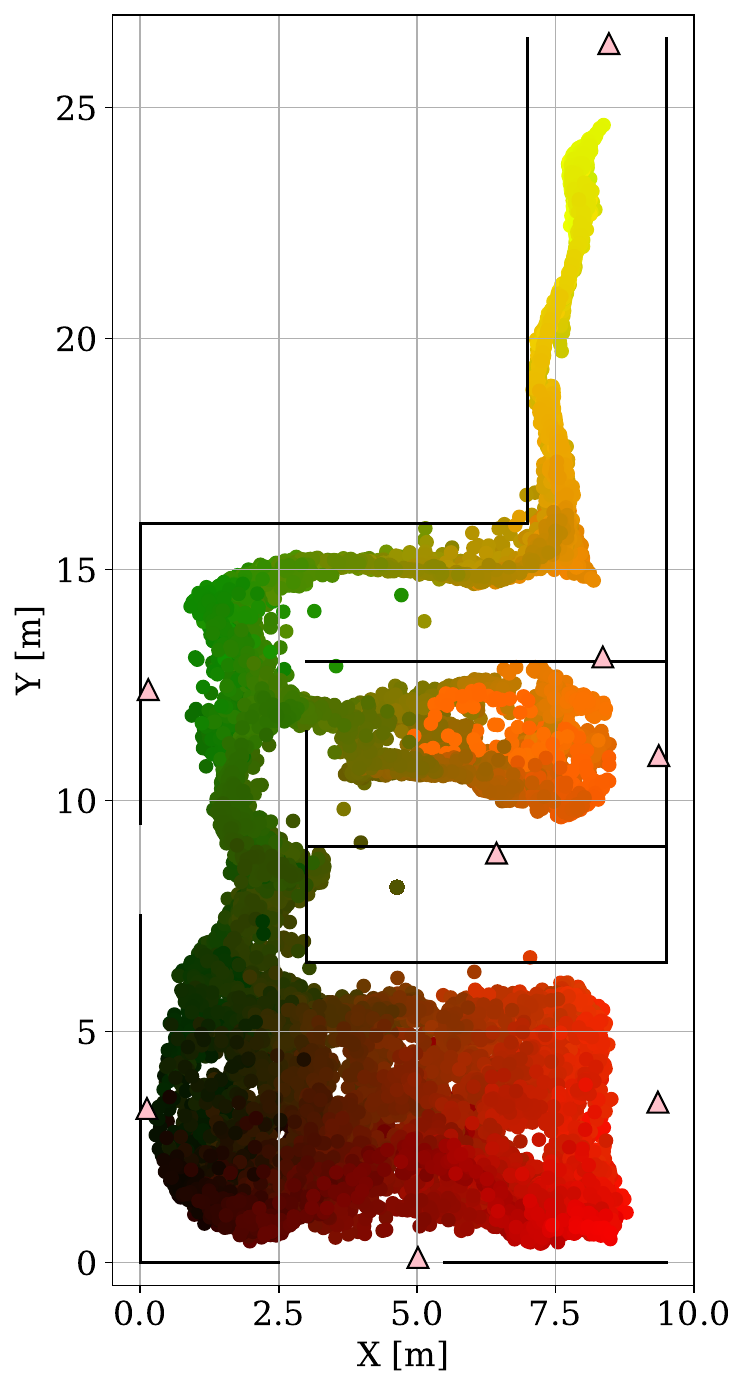}
        \caption{\textit{Power}.}
        \label{fig:pos_diff_height_pow}
    \end{subfigure}
    \begin{subfigure}{0.23\textwidth}
        \centering
        \includegraphics[width=\textwidth]{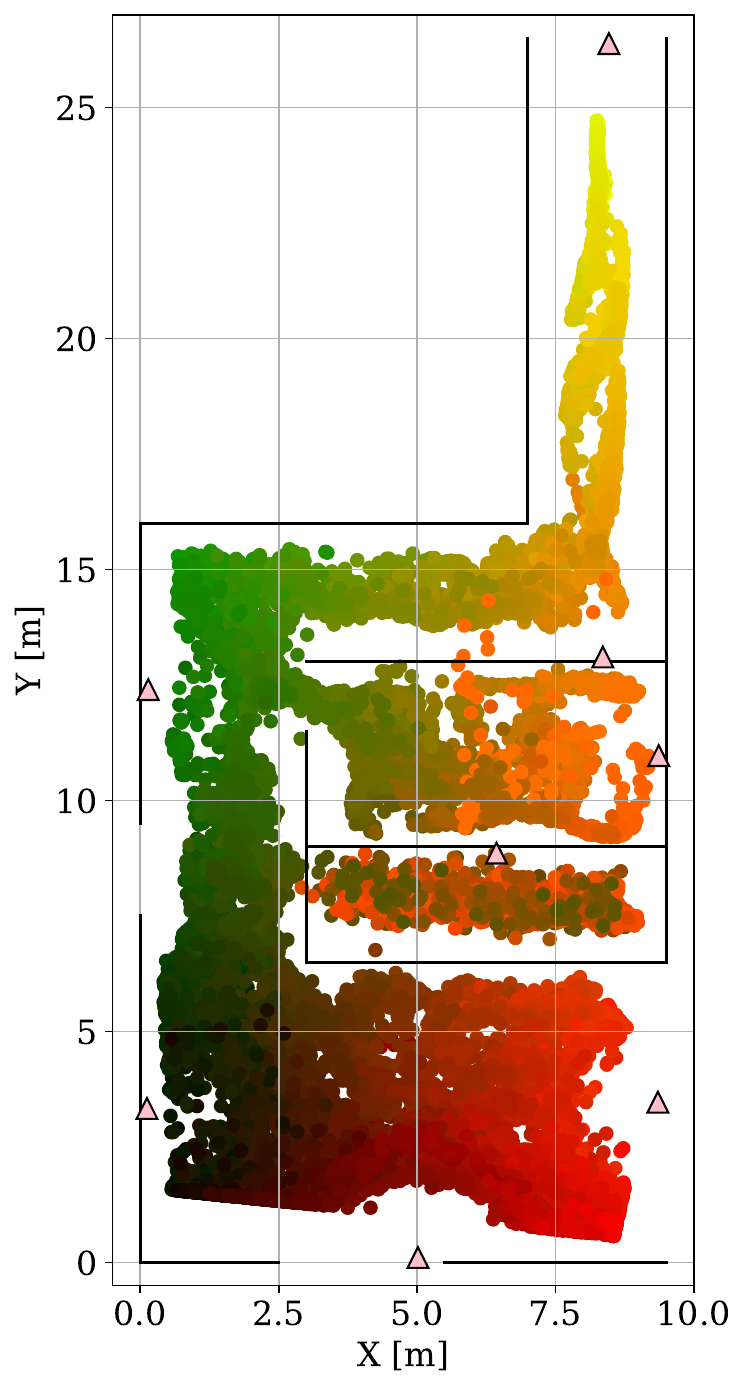}
        \caption{\textit{CSI fingerprinting.}}
        \label{fig:pos_diff_height_fingerprint}
    \end{subfigure}
    \caption{Estimated trajectories for the case where the UE height has been decreased during inference from $1.5\,$m to $0.8\,$m.}
    \label{fig:pos_diff_height}
\end{figure}

The performance slightly degrades under a UE height decrease compared to no mismatches in the DT. The degradation is mainly due to the pathloss effect, which we detail in the following. Fig.~\ref{fig:pos_diff_height} represents the estimated positions when the UE height is changed. 
Compared with Fig.~\ref{fig:pos_sim_results}, Fig.~\ref{fig:pos_diff_height} shows that no estimated positions appear near the APs (especially for the AP at the bottom center).
This occurs because pathloss primarily depends on the distance between the AP and the UE.
For example, the pathloss observed by a UE located at 1.5 m height and far from an AP in the horizontal plane is similar to that of a UE at 0.8 m height but closer in the horizontal plane.
As a result, when inference is performed at the lower height, positions that are actually close to the AP are misinterpreted as points located farther away in the horizontal plane.

\begin{figure}[tb]
    \centering
    % Top mini-figure (legend)
    \includegraphics[width=0.15\textwidth]{figures/results_sim/legend.pdf}\\[0.5ex]  % adjust the width and spacing here
    %Subfigures
    \begin{subfigure}{0.23\textwidth}
        \centering
        \includegraphics[width=\textwidth]{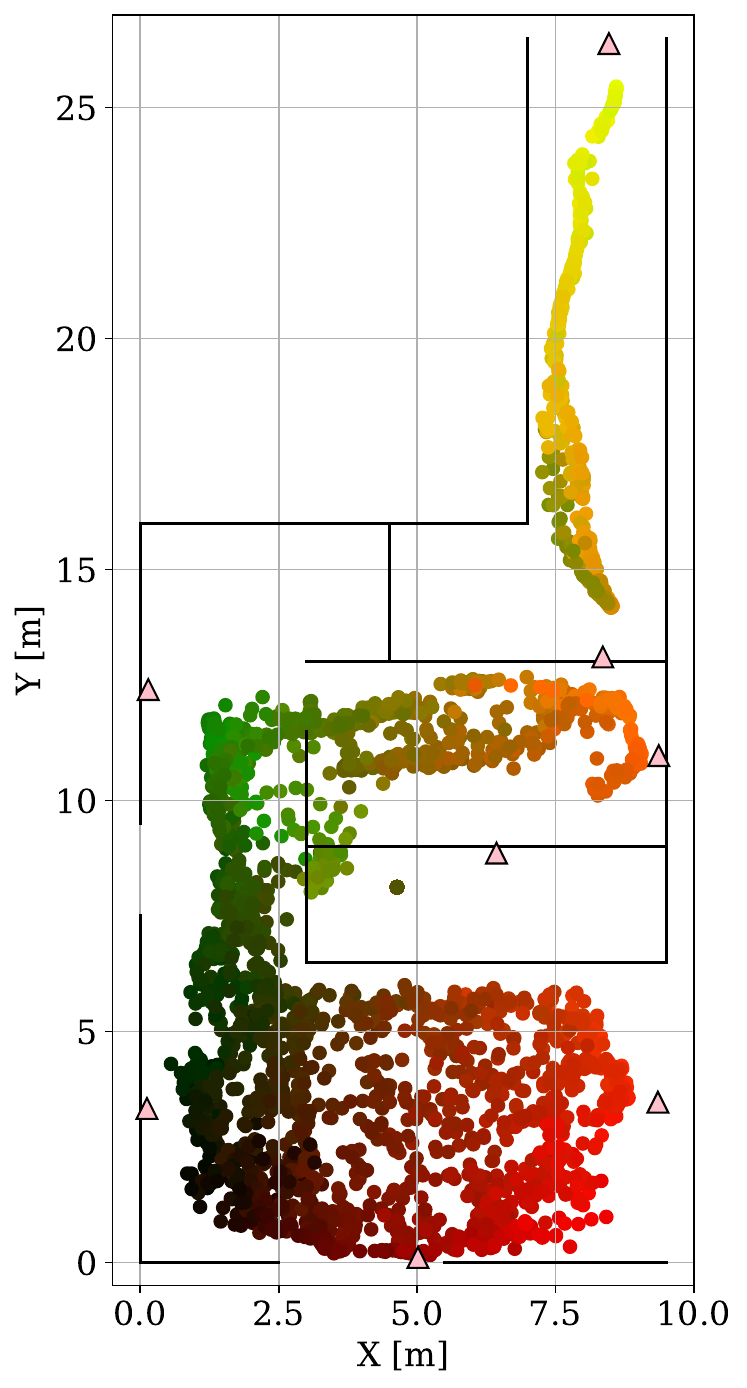}
        \caption{\textit{Power}.}
        \label{fig:results_door_pow}
    \end{subfigure}
    \begin{subfigure}{0.23\textwidth}
        \centering
        \includegraphics[width=\textwidth]{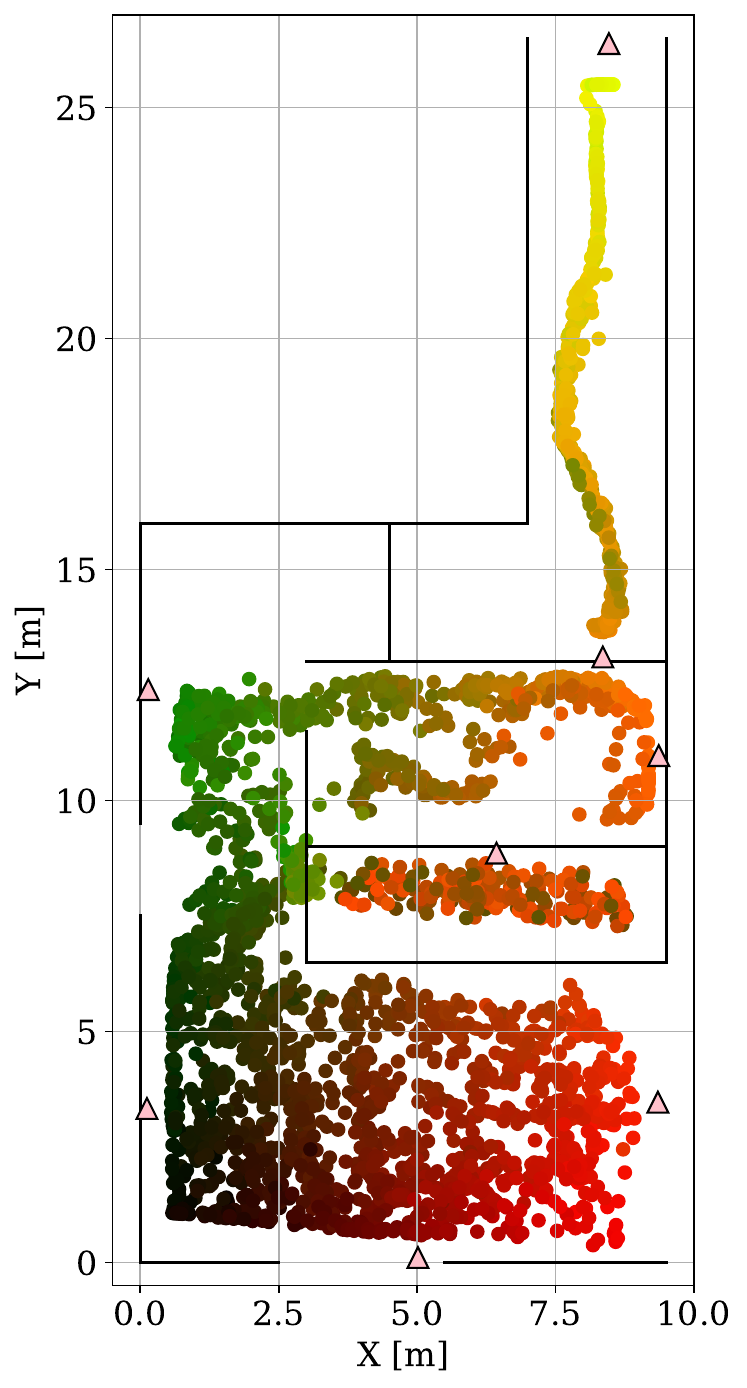}
        \caption{\textit{CSI fingerprinting.}}
        \label{fig:results_door_fingerprint}
    \end{subfigure}
    \caption{Estimated trajectories under a distribution shift of the data during testing time that includes a wooden wall in the top horizontal corridor.}
    \label{fig:results_door}
\end{figure}

On the other hand, introducing wooden boxes yields similar performance to matched testing data, which suggests that the large-scale features remain mostly unaltered. Conversely, adding a wall in the upper corridor degrades the performance of the proposed approach as it partially blocks the received signal of the APs in that region. The corresponding estimated trajectories are included in Fig.~\ref{fig:results_door}. Positions in the corridor near the added wall are estimated closer to the APs, indicating that the attenuated signals are interpreted as originating from more distant locations relative to the blocked AP.

The results in Table~\ref{tab:generalization} and Fig.~\ref{fig:pos_diff_height} indicate that the \textit{Power}, \textit{APP}, and \textit{DPP} features remain robust under distribution shifts that do not significantly alter the received signal at the APs.

\section{Conclusions and Outlook} \label{sec:conclusions}

In this work, we have proposed DT-aided CC for self-supervised positioning in an uplink D-MIMO scenario. The proposed framework preserves both the local and global geometry of the estimated positions without requiring labeled data, accurate AP synchronization, affine transformations, or LoS area estimation of each AP. Our framework leverages a DT to compute real-time large-scale feature vectors that help train conventional CC and provides global coordinates. We have investigated several feature vectors and showed that the \textit{Power}, \textit{APP}, and \textit{DPP} features outperform state-of-the-art results, with the \textit{Power} feature providing an overall better performance and the \textit{APP} and \textit{DPP} features more robustness in areas covered by only one AP. 
These results indicate that the proposed approach is an attractive alternative to CSI fingerprinting when a floor plan or DT of the environment is available.
We have also investigated the impact of the density of predefined DT positions, demonstrating that the \textit{Power} feature provides similar results with fewer computations from the DT.
Additionally, we have compared the performance of the proposed framework under DT modeling mismatches during training with two DT-based fingerprinting strategies that rely on synthetic CSI and received power, respectively. We concluded that, for slight mismatches, DT-based CSI fingerprinting achieves the best performance at the cost of increased computational complexity in the DT, while the proposed approach achieves the best performance when a computationally efficient DT is harnessed. Additionally, the proposed approach is the most robust under severe DT mismatches.
These results are the first step towards efficient, automatic, and real-time DT modeling, e.g., from pictures of the environment, without substantially degrading the positioning performance.

Finally, we have tested different datasets at inference time compared to the training data. We considered an unexpected decrease in the UE height or the introduction of reflecting elements in the environment. The results showcased that the proposed approach remains robust under distribution shifts that do not severely affect the large-scale features, still outperforming state-of-the-art methods. Under very severe distribution shifts, the proposed approach degrades in performance and fine-tuning with new collected unlabeled data must be performed to improve the overall positioning performance.

Accordingly, future research in this area can explore the 
continuous training of the proposed approach under changes in the environment and the
integration of tracking algorithms to further improve positioning performance while estimating the uncertainty of the predictions. Moreover, the proposed approach can be validated on real hardware with real measurements, where the NN can also account for hardware impairments.
Additionally, the proposed DT-based approach can be extended to other positioning modalities, e.g., matching features in camera images with simulated view scenes from the DT using ray-tracing algorithms.

\section*{Acknowledgments}
The authors would like to thank Remcom for providing a license for the Wireless InSite ray-tracing software. The authors also thank Sueda Taner for discussions on channel charting in real-world coordinates~\cite{taner2025channel}.

\appendix[Evaluation Metrics] \label{app:eval_metric}

In what follows, we list the metrics that we used to assess the positioning performance of the proposed method and the baselines. 
We denote by $\Jsetn$ and $\Jsethatn$ the sets of the $J$ closest true and estimated positions to $\xbm[n]$ and $\xbmhat[n]$, respectively. We denote the ranking of how close $\xbm[n]$ is to $\xbm[j]$ and how close $\xbmhat[n]$ is to $\xbmhat[j]$ according to the Euclidean distance as $r(n,j)$ and $\rhat(n,j)$, respectively. We define $\gamma=2/(NJ(2N-3J-1))$ as a normalization factor \cite{altous2022channel}. 
We set $J=0.05N$ as in \cite{studer2018channel}.

\subsubsection{Trustworthiness (TW)}
This metric penalizes close estimated positions that do not correspond to close true positions and it is defined as \cite{vathy2013graph}
\begin{align} \label{eq:tw}
    \TW(J) = 1-\gamma\sum_{n\in\Nset} \sum_{\substack{j\notin\Jsetn \\ j\in\Jsethatn}} (r(n,j)-J).
\end{align}
TW takes values in $[0,1]$ with optimal value $1$.

\subsubsection{Continuity (CT)}
This metric penalizes close true positions that do not correspond to close estimated positions and it is defined as \cite{vathy2013graph}
\begin{align} \label{eq:ct}
    \CT(J) = 1-\gamma \sum_{n\in\Nset} \sum_{\substack{j\in\Jsetn \\ j\notin\Jsethatn}} (\rhat(n,j)-J).
\end{align}
CT takes values in $[0,1]$ with optimal value $1$.

\subsubsection{Kruskal Stress (KS)}
This metric measures the dissimilarity between pairwise distances for true positions, defined as  $d(n,j)=\norm{\xbm[n]-\xbm[j]}$, and pairwise distances for the estimated positions, defined as $\dhat(n,j)=\norm{\xbmhat[n]-\xbmhat[j]}$. KS is defined as \cite{Kruskal_1964}
\begin{align}
    \KS = \min_{\eta\in\realset} \sqrt{\frac{\sum_{n,j\in\Nset} (\dhat(n,j)-\eta d(n,j))^2}{\sum_{n,j\in\Nset} \dhat(n,j)^2}}.
\end{align}
KS takes values in $[0,1]$ with optimal value $0$.

\subsubsection{Rajski Distance (RD)}
We denote as $V$ and $Q$ the discrete random variables representing the distribution of pairwise distances for the true positions $d(n,j)$ and the estimated positions $\dhat(n,j)$, respectively. The RD measures the difference between the mutual information and joint entropy of $V$ and $Q$ and it is defined as \cite{rajski1961metric}
\begin{align}
    \RD = 1-\frac{I(V,Q)}{H(V,Q)}, \text{for } H(V,Q)\neq 0,
\end{align}
where $I(V,Q)$ is the mutual information given by
\begin{align} \label{eq:mutual_info}
    I(V,Q)=\sum_{v\in V, q\in Q} P_{V,Q}(v,q)\log_2 \frac{P_{V,Q}(v,q)}{P_V(v)P_Q(q)}
\end{align}
and $H(Q,V)$ is the joint entropy given by
\begin{align} \label{eq:entropy}
    H(Q,V)=-\sum_{v\in V, q\in Q} P_{V,Q}(v,q) \log_2P_{V,Q}(v,q).
\end{align}
In \eqref{eq:mutual_info} and \eqref{eq:entropy}, $P_{V,Q}(v,q)$ is the joint distribution and $P_V(v)$ and $P_Q(q)$ are the marginal distributions of $V$ and $Q$, respectively. 
The RD takes values in $[0,1]$ with optimal value $0$. We quantize the pairwise distances for the true and estimated positions into 20 uniform bins.

\subsubsection{Mean Distance Error (MDE)}
This metric measures the average error between the true and estimated positions and it is defined as
\begin{align}
    \MDE = \frac{1}{N} \sum_{n\in\Nset} \norm{\xbm[n]-\xbmhat[n]}.
\end{align}
The MDE is nonnegative with optimal value $0$.

\subsubsection{95th Percentile Distance Error (PDE)}
This metric measures the 95th percentile of the norm between the true and estimated positions. It is defined as the value $\rho_{95}$ such that at least 95\% of the distance errors $\{ \norm{\xbm[n]-\xbmhat[n]} \}_{n\in\Nset}$ are less than $\rho_{95}$.

\bibliographystyle{IEEEtran}
\bibliography{references}

\balance

\end{document}